\documentclass[aps,prd,floatfix,nofootinbib,superscriptaddress,twocolumn,showkeys]{revtex4-2}

\usepackage{amsmath}%,amssymb}
\usepackage{listings}
\usepackage{amssymb}
\allowdisplaybreaks
\usepackage{float}
\usepackage{makeidx}
\usepackage{amsfonts}
\usepackage[usenames,dvipsnames]{pstricks}
\usepackage{subfigure}
\usepackage{epsfig}
\usepackage{pst-grad} % For gradients
\usepackage{mathtools}
\usepackage[colorlinks,hyperindex]{hyperref}
\usepackage{array}
\usepackage{color}
\hypersetup
{	colorlinks,%
	citecolor=black,%
	linkcolor=black,%
	urlcolor=black,%
}

\usepackage{allpurpose}

\usepackage{flushend}% Align the end of the last page.

\newcommand\pnp{{\text{PN}\Phi}}
\newcommand\pnu{{\text{PNU}}}
\newcommand\nn{{\nonumber}}

\def\XXint#1#2#3{{\setbox0=\hbox{$#1{#2#3}{\int}$}	\vcenter{\hbox{$#2#3$}}\kern-.5\wd0}}

\usepackage{listings}
\usepackage{ulem}
\usepackage{color}

\newcommand{\delred}[1]{{\color{red}{\ifmmode\text{\sout{\ensuremath{#1}}}\else\sout{#1}\fi}}}

\begin{document}

\title{Periapsis precession in general stationary and axisymmetric spacetimes}

\author{Jinhong He}
\thanks{These authors contributed equally to this work.}
\address{Department of Astronomy, School of Physics and Technology, Wuhan University, Wuhan, 430072, China}

\author{Shaofei Xu}
\thanks{These authors contributed equally to this work.}
\address{School of Physics and Technology, Wuhan University, Wuhan, 430072, China}

\author{Junji Jia}
\email[Corresponding author:~]{junjijia@whu.edu.cn}
\address{Department of Astronomy \& MOE Key Laboratory of Artificial Micro- and Nano-structures, School of Physics and Technology, Wuhan University, Wuhan, 430072, China}

\date{\today}

\begin{abstract}
This work studies the periapsis shift in the equatorial plane of arbitrary stationary and axisymmetric spacetimes. Two perturbative methods are systematically developed. The first works for small eccentricity but very general orbit size and the second, which is post-Newtonian and includes two variants, is more accurate for orbits of large size but allows general eccentricity.
Results from these methods are shown to be equivalent under small eccentricity and large size limits. The periapsis shift of Kerr-Newman, Kerr-Sen and Kerr-Taub-NUT spacetimes are computed to high orders. The electric charge and NUT charge are shown to contribute to the leading order but with opposite signs. The frame-dragging term and high-order effect of spacetime spin are given. The electric and NUT changes of the Earth, Sun and Sgr A* are constrained using the Mercury, satellite and S2 precession data. Periapsis shifts of other spacetimes are obtained too.

\end{abstract}


\maketitle

\section{Introduction}

The extra precession of Mercury's perihelion, or its periapsis shift (PS) around the Sun was one important observation made by Le Verrier, although his explanation was not correct \cite{lev1859}. It became one of the main observational foundations of General Relativity (GR) in its early days after being successfully explained by Einstein in 1916 \cite{einstein1916}.
With the progress of observation technology, the precession of periapsis of test particles' orbits has been discovered for many celestial bodies, including all planets in the solar system \cite{Iorio:2018adf}, among the binary star systems \cite{binary1988}, and in the center of the Galaxy around its supermassive black hole (BH) Sgr A* \cite{GRAVITY:2020gka}. Astronomers have identified and computed tens of the source for Mercury's PS \cite{park2017}, and multiple kinds of precession can now be separately observed.
With the development of space science, the precession of man-made objects can also be observed \cite{lucchesi2011,Everitt:2011hp}
and should be taken into account when designing orbits for satellites.

%Among these, the PS of satellites and S stars around the Sgr A* has drawn more attention lately. Experiments involving the former can be more precisely designed/controlled and therefore can be used to test gravitational theories to higher precision. Whereas the latter provides a special opportunity to test gravity in a stronger limit and can be related to other phenomena such as BH accretion.

Theoretically, the PS was computed mainly using two schemes.
The first is to use mainly the Newtonian mechanics with GR effects treated (at most) linearly as perturbation \cite{huda2006}. This approach is more useful when studying effects such as central object oblateness and third-object perturbation to the PS. The second approach starts from the equations of motion in GR but usually utilizes some kind of expansion when solving the complicated equation system. This will also be the approach taken by the present work. When such expansions are done to high orders, this approach can be considered as general relativistic. This method allows to find high-order post-Newtonian (PN) effects but is only limited to the cases where the spacetime is mainly determined by the central object and is usually highly symmetric.
Moreover, up to now, the second approach has only been used to study the PS in particular spacetimes. Ref. \cite{2006Precession,Shchigolev:2016vpz,Mak:2018hcn,Walter2018,Tucker:2019CQG} studied the PS in Schwarzschild spacetime,   Ref. \cite{Gong:2009zzb,A2011The,Hu:2013eya,Shchigolev:2016vpz,Mandal2022,Hall:2022ekh} in Reissner-Nordstr\"om (RN) spacetime, and Ref. \cite{Brill1997Light,Colistete:2002ka,Jiang2014} in Kerr spacetime and Ref. \cite{Heydari-Fard:2019kno} in Kerr-Newman (KN) spacetime. Gong and Wang studied the PS in the Kerr-Newman-Kasnya spacetime \cite{gong2008}.  Sutana et al. used the PS in a conformal Weyl gravity to constrain the parameter $\gamma$ there \cite{Sultana:2012qp}. Sun et al. used the PS of various objects to constrain the parameters of the fifth force \cite{Sun:2019ico}.  Zhang and Xie constrained the bounce parameter $\alpha$ using the PS of S2 in black-bounce-RN spacetime
\cite{Zhang:2022zox}.
Results in many of these works are expressed in terms of intermediate variables (e.g. the angular momentum $L$) \cite{A2011The,Hu:2013eya,Jiang2014,Shchigolev:2015sgg} that are not very convenient to compare with observational data. Some other works are only limited to the circular orbit limit \cite{Colistete:2002ka,LeTiec:2013uey,Hinderer:2013uwa,Heydari-Fard:2019kno,Bakry:2021smq}.

The above observations motivate us, in this work, to develop two methods that can find the PS in the equatorial plane of arbitrary stationary and axisymmetric (SAS) spacetimes. The first method uses essentially a small eccentricity expansion of the orbits and applies to orbits with very small radius. The second method expands the PS to high orders of the orbit size and therefore is more accurate in the weak field limit. The second method has two variants which are shown to be equivalent. When both the small eccentricity and large orbit size conditions are satisfied, the two methods are shown to yield the same result for general SAS spacetimes too.
However, the exclusive application of results from the first method in astronomy/observation is very limited due to the rareness of objects with small orbital radii (e.g., those near the innermost stable circular orbits) but still can be treated as test particles influenced by the central body only. Therefore in Sec. \ref{sec:appinastro}, only results using large radius expansion are applied to several spacetimes, for which we used the Mercury, satellite and S2 data to constrain the electric charge of the Sun, Earth, and Sgr A*, as well as their NUT charge.

The paper is organized as follows. In Sec. \ref{sec:pertmeth}  we first present the preliminaries, and then fully develop the two methods, and show their equivalence under certain conditions. In Sec. \ref{sec:appinspacetimes}, the results are then applied to the KN, Kerr-Sen (KS) and Kerr-Taub-NUT (Newman-Unti-Tamburino) (KTN) spacetimes, whose results are used to show the correctness of our method. We also carefully analyze the effects of various parameters influencing the PS.
Sec. \ref{sec:appinastro} is devoted to the constraining of the spacetime parameters using observational data of precession. We conclude the paper with Sec. \ref{sec:concdis}. Through out the paper, we use the natural units $G=c=1/(4\pi\epsilon_0)=1$ and signature $(-,\,+,\,+,\,+)$.

\section{The perturbative methods\label{sec:pertmeth}}

In this section, we present two perturbative methods that work for general SAS spacetime to compute the PS.
The first is a perturbative method, based on a change of variables that we developed previously \cite{Jia:2020qzt} for the computation of the deflection angles of the trajectory. This method is entirely general relativistic and works even when the orbit is in the strong field limit, i.e., close to the black hole. However, this method is most precise for near-circular orbits.
The second method is a PN one and works for orbits with larger eccentricity but is more accurate only when the orbit is far from the center.

We will study the precession in the equatorial plane of general SAS spacetimes, which of course, include static and spherically symmetric spacetimes.
In Boyer-Linquidist coordinates $(t,\,r,\,\theta,\,\phi)$, an arbitrary SAS spacetime can be described by the line element
        \begin{equation}\label{eq:metric}
        \mathrm{d}s^2=-A(r)\mathrm{d}t^2+B(r)\mathrm{d}t\mathrm{d}\phi+C(r)\dd\Omega^2+D(r)\mathrm{d}r^2
        \end{equation}
where $\dd \Omega^2=\dd\theta^2+\sin^2\theta\mathrm{d}\phi^2$ and $A,\,B,\,C,\,D$ are functions of $r$ only.
The equation of motion of a test particle in the equatorial plane then is given by the following equations
\begin{subequations}
    \begin{align}
        \Dot{t}=&\frac{2(2EC+LB)}{4AC+B^2},\label{eq: geo1}\\
        \Dot{\phi}=&\frac{2(2LA-EB)}{4AC+B^2},\label{eq: geo2}\\
        \Dot{r}^2=&\frac{(4 A C + B^2) (E^2-\kappa A) - (2LA-EB)^2 }{AD (4AC+B^2)}\label{eq: geo3},
    \end{align}
\end{subequations}
where $\dot{~}$ stands for derivative with respect to the affine parameter and $\kappa=1$ for timelike test particle, $E$ and $L$ are the conserved energy and angular momentum per unit mass of the particle. Using the last two equations, we find
\begin{align}
\label{eq:dphiodrdef}
\frac{\dd \phi}{\dd r}=&\sqrt{\frac{AD}{AC+B^2/4}}\nn\\
            &\times \frac{(2L A-E B)}{\sqrt{(4A C+B^2)\left(E^2-\kappa A\right)-(2L A-E B)^2}}
\end{align}
For later reference, we can also write this as an equation of $u\equiv 1/r$
\begin{align}
    &\left(\frac{\dd u}{\dd\phi}\right)^2= \frac{(4AC+B^2)u^4}{AD(4LA-2EB)^2}\nn\\
    &\times\left[(4 A C + B^2) (E^2-\kappa A) - (2LA-EB)^2\right],\label{eq: du/dphi}
\end{align}
where the metric functions here should be viewed as functions of $u$ too.

Since we are interested in the PS, we will only consider orbits bounded by a maximum radius and a minimum radius, which will be denoted as $r_1$ and $r_2$ respectively. By definition, $r_1,\,r_2$ are the solutions to the equation obtained by setting $\dot{r}\big|_{r=r_{1,2}}=0$. Using Eq. \eqref{eq: geo3}, we have
\begin{align}
(4 A C + B^2) (E^2-\kappa A) - (2LA-EB)^2 \Big|_{r=r_{1,2}}=0.
\end{align}
The angular momentum $L$ can be expressed in terms of the metric functions at the above $r_{1,2}$. Using Eq. \eqref{eq: geo3}, we can first solve
\begin{equation}
    L=\frac{EB}{2A}+s \frac{\sqrt{4AC+B^2}}{2A}\sqrt{E^2-\kappa A -AD \dot{r}^2},\label{eq:Linr}
\end{equation}
where $s=\pm 1$ is the sign produced when taking the square root. The first term is the contribution of the spacetime spin to the particle's total angular momentum and the second term represents the contribution from linear velocity. Substituting $r=r_{1,2}$, we immediately have
\begin{align}
    L=\frac{EB}{2A}+s \frac{\sqrt{4AC+B^2}}{2A}\sqrt{E^2-\kappa A}\bigg|_{r=r_{1,2}}.\label{eq:Linr12}
\end{align}

The PS, which will be denoted as $\delta$, then is defined as twice the angle swept by the orbit from $r_2$ to $r_1$ minus $2\pi$, i.e.,
\begin{equation}
    \delta = 2 \left(\int_{r_2}^{r_1}\frac{\dd \phi}{\dd r}\dd r-\pi\right),\label{eq:deltadef}
\end{equation}
where $\dd\phi/\dd r$ is given in Eq. \eqref{eq:dphiodrdef}. Later on when we substitute Eq. \eqref{eq:Linr12} into Eq. \eqref{eq:dphiodrdef} and then into Eq. \eqref{eq:deltadef}, we will use, without losing any generality, $s=1$ and therefore $\dd\phi/\dd r>0$ and the test particle rotates counterclockwise. This integral for known spacetimes in general however can not be carried out to find closed forms in terms of simple functions. One additional difficulty is that usually the precession is not expressed in terms of the minimal and maximal radii in the integral limits, but more characteristic variables of the orbit, such as the conserved $(L,\,E)$ or in the Newtonian limit the length of the semilatus rectum $p$ or semimajor axis $a$ and the eccentricity $e$. Therefore, one of our tasks in this paper is to develop perturbative methods to approximate this precession using some more common quantities.
More importantly, we will try to develop general methods that are not limited to specific spacetimes, but that can be applied to arbitrary SAS spacetimes in general.

As stated before, next we will present two methods: the first is a fully general relativistic treatment and is more accurate when the orbit is near a circle, i.e., the eccentricity is small.  The second method uses a PN ansatz and can be further split into two variants, according to what function in the ansatz is expanded. We will term the first method the small eccentricity (``SE'') method, and the two variants of the second method the ``PN$\Phi$'' and ``PNU'' methods.

\subsection{The SE method\label{subsec:grse}}

The first method is inspired by a transformation that we discovered in finding the deflection angle of an orbit in the strong deflection limit \cite{Jia:2020qzt}. Its main advantage is that it can be applied to arbitrary SAS spacetime and naturally express the result in terms of the variables of the orbit such as $(L,\,E)$.

To use this method, we first define a function $l(r)$
\begin{align}
    l(r)=\frac{E B(r)}{2A(r)}+\sqrt{\frac{B(r)^2}{4A(r)}+C(r)}\sqrt{\frac{E^2}{A(r)}-\kappa }.
    \label{eq:lrdef}
\end{align}
We will assume that $l(r)$ admits at least one local extreme value $l_c$ at $r=r_c$
\begin{align}
    \frac{\dd l(r)}{\dd r}\bigg|_{r=r_c}=0,~l(r_c)=l_c.
    \label{eq:dlreq}
\end{align}
From the definition \eqref{eq:lrdef} it is seen that this $r_c$ and $l_c$ can always be fixed once the metric functions and the energy $E$ of the particle are known, although their exact values can only be determined after the exact form of these functions or constants are chosen.
Comparing to $L$ in Eq. \eqref{eq:Linr}, we see that as long as we take a value of $L$ that is close to $l_c$ from an appropriate direction, then there will always exist some points near $r_c$ that satisfies $\dot{r}=0$. We let the solution to this equation be the $r_1$ and $r_2$ mentioned before. Naturally, we would have $r_2<r_c<r_1$ and the integral in the PS \eqref{eq:deltadef} can be split into two parts
\begin{equation}
\delta_{\text{SE}}=2\left[   \left(\int_{r_2}^{r_c}+\int_{r_c}^{r_1}\right)\frac{\dd \phi}{\dd r}\dd r -\pi\right].\label{eq:delta2p}
\end{equation}
We will refer to the ranges of $r$ from $r_2$ to $r_c$ and $r_c$ to $r_1$ as part two and part one respectively.

To proceed, we still need to define a function $h(x)$, with the help of $l(r)$, as
\begin{equation}
    h\left(\frac{1}{r}\right)=\frac{l_c-l(r)}{L}. \label{eq:hrdef}
\end{equation}
Then for $r$ in part one, we can find the inverse function of $h(x)$ and denote it as $q_1(x)$. Similarly for $r$ in part two, we can find the inverse function $q_2(x)$. We point out that once the metric functions are given and $l(r)$ in Eq. \eqref{eq:lrdef} is defined, it is always possible to establish such inverse maps. Although the explicit and closed forms of these inverse functions might be difficult to obtain, it is also known that these functions can be obtained perturbatively through the Lagrangian inversion theorem, and these are exactly what is needed later for our SE method. Moreover, since these inverse functions are on different sides of an extreme point of the original function $h(x)$, in principle there can exist two different branches, which we denoted as $q_1(x)$ and $q_2(x)$.

The key step to process is then for  part $i~(i=1,\,2)$ of the integral, we use a change of variables from $r$ to $\xi$ linked by the relation \eqref{eq:hrdef} or its inverse function, i.e.,
\begin{align}
    h\lb \frac1r\rb=\xi,~r=\frac{1}{q_i(\xi)}. \label{eq:rtoxi}
\end{align}
Using this and Eq. \eqref{eq:dphiodrdef}, the limits and integrands in the precession \eqref{eq:delta2p} can be transformed into
\begin{subequations}
\begin{align}
r\to& \frac{1}{q_i(\xi)},\\
r_i\to& \frac{l_c}{L}-1\equiv \eta,\label{eq:etadef}\\
r_c\to& 0,\\
\frac{\dd \phi}{\dd r}\to& \sqrt{\frac{A(1/q_i(\xi))D(1/q_i(\xi))}{A(1/q_i(\xi))C(1/q_i(\xi))+B(1/q_i(\xi))^2/4}}\nn\\
&\times \frac{\beta(1/q_1(\xi))}{\sqrt{(\eta-\xi+2\beta(1/q_1(\xi))}\sqrt{\eta-\xi}},
\end{align}
\end{subequations}
where we have defined the value
\begin{equation}
l_c/L-1\equiv \eta ~\text{and}~\beta(r) = 1-\frac{E B(r)
}{2 L A(r)}.
\end{equation}
The precession \eqref{eq:delta2p} then becomes
\begin{align}
    \delta_{\text{SE}} =&2\int_0^{\eta}\dd \xi \frac{1}{\sqrt{\eta-\xi}}\sum_{i=1,2}(-1)^{i+1}\left(-\frac{q^\prime_i(\xi)}{q_i(\xi)^2}\right)\nn\\
    &\times \sqrt{\frac{AD}{AC+B^2/4}} \frac{\beta}{\sqrt{\eta-\xi+2\beta}}-2\pi \label{eq:dinsum}\\
&\equiv 2\int_0^{\eta}\frac{Y(\xi)}{\sqrt{\eta-\xi}}\dd\xi-2\pi,
\label{eq:qphitransed}
\end{align}
where the metric functions and $\beta$ in Eq. \eqref{eq:dinsum} are evaluated at $1/q_i(\xi)$ and in Eq. \eqref{eq:qphitransed} the entire summation is defined as $Y(\xi)$.

The above change of variables allows us to transform the limits $r_{1,2}$ to the variable $\eta$ which is linked to more fundamental variables $l_c$ and $L$. More importantly, if $\eta$ is a small quantity, on which case we will focus in this method, we can expand $Y(\xi)$ for small $\xi$ and show that the resulting integrals can always be carried out to yield a series form. The accuracy of this series can be achieved to arbitrarily high as long as the expansion is done to a high enough order. To be more precise, expanding $Y(\xi)$, we can find that it always takes a form (see also the Appendix \ref{app:yhalfzero})
\begin{align}
    Y(\xi)=\sum_{n=0}y_n \xi^{n-1/2},\label{eq:yexp}
\end{align}
where $y_n$ are the expansion coefficients and can be obtained immediately once the metric functions and $(L,\,E)$ are known. To show this, let us assume that the metric functions have the following expansion at $r_c$
\begin{subequations}
\label{eq:abcdexpatrc}
    \begin{align}
         A(r)=&\sum_{i=0}^\infty a_i\lb \frac{1}{r}-\frac{1}{r_c}\rb^i,\\
        B(r)=&\sum_{i=0}^\infty b_i\lb \frac{1}{r}-\frac{1}{r_c}\rb^i,\\
        C(r)=&\sum_{i=0}^\infty c_i\lb \frac{1}{r}-\frac{1}{r_c}\rb^i,\\
        D(r)=&\sum_{i=0}^\infty d_i\lb \frac{1}{r}-\frac{1}{r_c}\rb^i,
    \end{align}
\end{subequations}
where $a_i, b_i, c_i, d_i$ are expansion coefficients that contain $r_c$. Here we used an expansion of $1/r$ at $1/r_c$ rather than $r$ at $r_c$ because for some known spacetime such as Schwarzschild, Kerr and KN, some metric functions become a finite sum in the former expansion. Using these expansions in Eqs. \eqref{eq:hrdef} and then finding the inverse functions $q_i(x)$ and further using Eq. \eqref{eq:qphitransed}, we are able to work out the coefficients $y_n$.
Since we try to be as general as possible in our derivation, these $y_n$ are usually too long to present here, except for the first few
\begin{widetext}

    \begin{align}
    &y_0=\left[\frac{2 a_0 L-b_0 E+a_0 L \eta}{a_0 d_0 r_c^4 \left(2 a_0 L-b_0 E\right)^2}\right]^{-1/2}\left\{ \left(a_1^2 b_0 c_0+a_0 a_1 b_0 c_1+\frac{1}{4} a_1^2 b_0^2+a_1^2 c_0^2+2 a_0 a_1 c_0 c_1+a_0^2 c_1^2\right)\frac{v}{\Delta_0}\right. \nn
    \\
    &+  \left[\left(\frac{1}{4} a_1^2 b_0-\frac{a_1^2 b_0^2}{4 a_0}-\frac{1}{2} a_1^2 c_0+\frac{1}{2} a_0 a_1 c_1\right)\frac{1}{v}+ \left(\frac{a_1^2 b_0}{2 a_0}+\frac{a_1^2 c_0}{a_0}-a_2 c_0-a_0 c_2-\frac{a_1^2}{4}-\frac{b_0 b_2}{2}\right)v\right]\Delta_0
\nn\\
    &\left. \left(-\frac{a_1^2 b_0}{2 a_0^2}+\frac{a_2 b_0}{2 a_0}+\frac{a_1^2}{2 a_0}-\frac{b_2}{2}\right)E\Delta_0 ^2+ \left[\frac{a_1^2}{16 v^3}+\frac{a_2}{4 v}+\left(\frac{a_2}{2 a_0}-\frac{a_1^2}{2 a_0^2}\right) v\right]\Delta_0 ^3\right\}^{-1/2},
    \end{align}
\end{widetext}
where $v^2\equiv E^2-a_0, \Delta_0 ^2\equiv 4 a_0 c_0+b_0^2$.

Substituting the expansion \eqref{eq:yexp} into Eq. \eqref{eq:qphitransed}, the precession is converted to a series of integrals of the form
\begin{equation}
    \delta_{\text{SE}} =2\sum_{n=0}^{\infty}y_{n}\int_0^{\eta}\frac{\xi^{n-1/2}}{\sqrt{\eta-\xi}}-2\pi.
\end{equation}
Fortunately, the integrals of the above form can always be carried out to yield integer power functions of $\eta$
\begin{align}
I_n=\int_0^{\eta}\frac{\xi^{n-1/2}}{\sqrt{\eta-\xi}}\dd \xi=\frac{\pi(2n-1)!!}{(2n)!!}\eta^n,~~n\in\mathbb{Z}_\geq .
\end{align}
This leads to the SP of the final form
\begin{align}
\delta_{\text{SE}} =&\sum_{n=0}^{\infty}
    \frac{2\pi(2n-1)!!}{(2n)!!}y_n\eta^n
-2\pi \label{eq:deltaf1}\\
=&2\pi(y_0-1)+\pi y_1\eta+\frac34 y_2\eta^2+\mathcal{O}(\eta)^3,
    \label{eq:dintheta2}
\end{align}
where in the last equation we used the first few $I_n$ and showed the leading orders of $\delta$.

A few comments about this result are in order. The first feature of this PS is that it is nonzero even in the circular orbit limit $\eta\to 0$. In other words, $y_0\neq 1$ in general.
Moreover, we also point out that for Kerr-like spacetimes whose metric function $B(r)\propto\hat{a}$, the effect of the $\hat{a}$ generally will show up even in the first order of $\delta$, i.e., in $y_0$. This is in contrast to the effect of $\hat{a}$ on the trajectory deflection angle, which only appears in orders equal to or higher than two \cite{Huang:2020trl,Liu:2020mkf}. Thirdly we note that Eq. \eqref{eq:dintheta2} is only a quasi-power series of $\eta$,  because the coefficients $y_n$ also contain weak $\eta$ tdependence.
Finally, we emphasize that this result effectively is a small $\eta$ expansion and therefore is more accurate if the orbit is near circular, which is unlike the PN$\Phi$ or PNU methods in Subsec. \ref{subsec:pn} which are applicable for quite general $e$. However, the result here has a great advantage in that it is valid in the strong gravity regime, i.e., when the orbit is very close to the BH center.

\subsection{The PN methods\label{subsec:pn}}

As pointed out in the introduction section, the PN methods for solving the PS have been used extensively.
In this work, we will systematize two PN methods to arbitrary SAS spacetime, paying special attention to the convergence and higher orders of the series result.
Both methods start by assuming that the solution to the orbit can be approximated to the lowest order by a Keplerian elliptic form
\begin{align}
    u_0\equiv\frac1r
=\frac{1}{p}+\frac{e\cos{\phi}}{p},
\label{eq:anzorder0}
\end{align}
where $p$ is the semiletus rectum and $e$ is the eccentricity.
One of the methods then generalizes the $\phi$ to a function $\psi=\psi(\phi)$ and attempts to find a perturbative solution of this function to account for the effect of the GR to the orbital equation, and therefore is termed the PN$\Phi$ method. The other method, which is inspired by the Poincar\'{e}-Lighthill-Kuo method \cite{plkmethod},
besides modifying $\phi$, also generalizes the $u_0$ on the left-hand side to  $u=u_0(\phi)+u_1(\phi)/p+\cdots$, i.e. a series of functions $u_i(\phi)~(i=0,\,1,\,2,\,\cdots)$, with $u_0(\phi)$ given by the right-hand side of Eq. \eqref{eq:anzorder0}. Therefore we term this method as the PNU method.

In the following, we present these methods in detail. As we will see, both methods actually use the large orbital radius limit, which is the fundamental reason that both methods are said to be PN.

\subsubsection{The PN\texorpdfstring{$\Phi$}. method \label{subsubsec:pnone}}

In this method, we first propose the following ansatz for the solution of Eq. \eqref{eq: du/dphi}
\begin{align}
    u\equiv\frac1r
=\frac{1}{p}+\frac{e\cos{\psi(\phi)}}{p},
\label{eq:anz1}
\end{align}
where $\psi(\phi)$ is a function to be determined and $p,\,e$ are to be connected with constants of the motion $L$ and $E$. Compared to the Keplarian solution, this basically is a generalization of $\phi$ to $\psi(\phi)$ but the left-hand side, $u$, is still directly interpreted as the inverse of $r$.
At the minimal radius $r_2$ and maximal radius $r_1$ of the radial motion, then clearly  $\cos\psi(\phi)$ should take its extremes, i.e., $\psi=0$ and $\psi=\pi$ respectively. Besides, it is obvious that at these boundaries, we also have
\begin{align}
\lb \frac{\dd u}{\dd \psi}\rb^2\Bigg|_{\psi=0}=0,~\lb \frac{\dd u}{\dd \psi}\rb^2\Bigg|_{\psi=\pi}=0.
\label{eq:dudpsibd}
\end{align}

To determine $\psi(\phi)$, we still need to work with Eq. \eqref{eq: du/dphi}. Denoting the right-hand side of this equation as $F(u)$, i.e.,
\begin{align}
F(u)=& \left[(E^2-\kappa A)(B^2+4AC)-(2LA-EB)^2\right]\nn\\
    &\times\frac{(4AC+B^2)u^4}{AD(4LA-2EB)^2},\label{eq:fudef}
\end{align}
and appealing to the large radius approximation or equivalently the PN limit, so that $u$ is a small quantity for the entire orbit, we can expand $F(u)$ at $u=0$. Thus Eq. \eqref{eq: du/dphi} becomes
\begin{equation}
 \left(\frac{\dd u}{\dd\phi}\right)^2=F(u)=\sum_{n=0}^{\infty}f_n u^n,\label{eq:fuexpd}
 \end{equation}
where $f_n$ are the coefficients independent of $u$ and $\phi$ but dependent on $E,\,L$. The exact forms of $f_n$ can be linked easily to the metric functions. Assuming they have the following asymptotic expansions
\begin{align}
    &A(r)=\sum_{n=0} \frac{a_n}{r^n},~B(r)=\sum_{n=1} \frac{b_n}{r^n},\nn\\
    &C(r)=\sum_{n=0} \frac{c_n}{r^{n-2}},~D(r)=\sum_{n=0} \frac{d_n}{r^n},
    \label{eq:metricexp}
\end{align}
then since PN methods only work in asymptotically flat spacetimes, we immediately see that $a_0=c_0=d_0=1$ should be required. Substituting the above expansion into Eq. \eqref{eq:fudef}, we can work out all $f_n$'s, the first few of which can be shown explicitly as
\begin{subequations}
\label{eq:fexpcoeff}
    \begin{align}
        f_0=&\frac{E ^2-1}{  L^2},\\
        f_1=&\frac{E^2-1}{L^2}\lb \frac{b_1 E}{ L}+2c_1- d_1\rb-\frac{a_1 E^2}{L^2},\\
        f_2=&-1+\frac{1}{L^2}\left\{\left[a_1 \left(a_1-2c_1+d_1\right)-a_2\right]E^2\right.\nn\\
        &\left.+\left[(d_1-c_1)^2+2 c_2-d_2\right](E^2-1) \right\}\nn\\
        &+\frac{E}{L^3}\left[a_1 b_1\left(1-2E ^2\right)+\left(E^2-1\right) \left(b_2+2b_1c_1-b_1 d_1\right)\right]\nn\\
        &+\frac{3 b_1^2 E^2 \left(E^2-1\right)}{4 L^4},
    \end{align}
\end{subequations}
while the higher orders can be similarly obtained.

Now substituting the ansatz \eqref{eq:anz1} into the right-hand side of Eq. \eqref{eq:fuexpd}, it becomes
 \begin{align}
 \left(\frac{\dd u}{\dd\phi}\right)^2=\sum_{n=0}^{\infty}\frac{f_n}{p^n}\lb 1+e\cos\psi\rb^n .\label{eq:fuexpdsub}
 \end{align}
Further expanding the binomial and using
\begin{align}
    \cos^{2m}\psi=&(1-\sin^2\psi)^m\nn\\
    =&1-\sin^2\psi\sum_{k=1}^mC_m^k(\cos^2\psi-1)^k,\nn\\
\cos^{2m+1}\psi=&\cos\psi-\sin^2\psi\sum_{k=1}^mC_m^k(\cos^2\psi-1)^k \cos\psi,\nn
\end{align}
for positive integer $m$, we can collect the right side into a sum of three functions
\begin{equation}
     \lb \frac{\dd u}{\dd \phi}\rb^2=F_1+F_c \cos\psi
     +F_s(\cos\psi)\sin^2\psi,
     \label{eq:dudpsisq}
 \end{equation}
where $F_1,\,F_c,\,F_s$ are all linear combinations of $f_n$ with coefficient functions to be integer power series of $1/p$ and $e$. Moreover, $F_c$ and $F_s$ are apparently also integer power series of $\cos\psi$. All the exact forms of these three functions can be worked out but we first only show $F_s$ here \begin{align}
    F_s(f_n,\,p,\,e,\cos\psi)=\sum_{n=0}\frac{s_n}{p^n}f_n,
    \label{eq:fsform}
\end{align}
with coefficients $s_n$ given by
\begin{align}
s_n=& -\Bigg[ \sum_{m=0}^{\left[\frac{n}{2}\right]}C_{n}^{2m} e^{2m}+\cos\psi \sum_{m=0}^{\left[\frac{n-1}{2}\right]} C_{n}^{2m+1} e^{2m+1}\Bigg]\nn\\
&\times \sum_{l=1}^{m}C_{m}^{l}(\cos^2\psi-1)^{l-1}.
\end{align}
It is easy to verify that
\begin{align}
s_0=s_1=0,~s_2=-e^2 \label{eq:sn02res}
\end{align}
and $s_n~(n\geq 3)$ are $n-2$ order polynomials of $\cos(\psi)$. It follows then the $f_0$ and $f_1$ coefficients of $F(u)$ do not contribute to the function $F_s$. For $F_1$ and $F_c$,
substituting Eq. \eqref{eq:dudpsisq} into \eqref{eq:dudpsibd}, we immediately see that
\begin{align}
&\lb \frac{\dd u}{\dd \phi} \rb^2\Bigg|_{\psi=0}=F_1+F_c=0,\nn\\
&\lb \frac{\dd u}{\dd \phi} \rb^2\Bigg|_{\psi=\pi}=F_1-F_c=0,
\end{align}
which lead to $F_1=F_c=0$.

Finally, we can establish a simple equation for $\psi(\phi)$ by substituting ansatz \eqref{eq:anzorder0} into the left-hand side of Eq. \eqref{eq:dudpsisq} and obtain
 \begin{equation}
     \frac{e^2}{p^2}\sin^2\psi\left(\frac{\dd \psi}{\dd \phi}\right)^2=\sin^2\psi F_s(\cos\psi). \label{eq:dpsidphisq}
 \end{equation}
Using this, the PS in Eq. \eqref{eq:deltadef} can then be converted to an integral over $\psi$ from $0$ to $\pi$,
\begin{align}
     \delta_\pnp =2\int_{\phi(r_1)}^{\phi(r_2)}\dd\phi -2\pi=2\int_0^{\pi}\frac{e}{p}\frac{1}{\sqrt{F_s(\cos\psi)}}\dd\psi-2\pi.
     \label{eq:psinpsi}
\end{align}

Since $F_s(\cos\psi)$ as given in Eq. \eqref{eq:fsform} is naturally a negative integer power series of $p$, we can also expand the integrand of \eqref{eq:psinpsi} in the large $p$ limit and then approximate $\delta_\pnp$ as a series of $1/p$ too. The very fact that $F_s$ is a positive integer power series of $\cos(\psi)$ also implies that each term of the integrand expansion is a finite order polynomial of $\cos(\psi)$, which guarantees the integrability of the expanded integrand. To be more specific, the expansion of the integrand of Eq. \eqref{eq:psinpsi} is found as
\begin{align}
\frac{e}{p}\frac{1}{\sqrt{F_s(\cos\psi)}}=\sum_{n=0}\frac{p_n(\cos\psi)}{p^n},
     \label{eq:psiintgdexp}
\end{align}
where $p_n$ are the coefficient polynomials, of which the first few can be explicitly written out
\begin{subequations}
\begin{align}
&p_0(\cos\psi)=\frac{1}{(-f_2)^{1/2}},\\
&p_1(\cos\psi)=\frac{-s_3f_3}{2e^2(-f_2)^{3/2}},\\
&p_2(\cos\psi)=\frac{3s_3^2f_3^2+4e^2s_4f_2f_4}{8e^4(-f_2)^{5/2}}.
\end{align}
\end{subequations}
Note that $p_n$ for arbitrary $n=0,\,1,\,\cdots$ is an $n$-th order polynomial of $\cos\psi$ and depends on $f_i$ only to the $(n+2)$-th order.
Denoting integral of $p_n$ as $P_n$, i.e.,
\begin{align}
    \int_0^\pi p_n(\cos\psi,f_2,\cdots,f_{n+2})\equiv &P_n(e,f_2,\cdots,f_{n+2}),
\end{align}
the $\delta_\pnp$ finally becomes
\begin{align}
    \delta_\pnp=2\sum_{n=0}\frac{P_n}{p^n}-2\pi, \label{eq:deltaf2}
\end{align}
with first few $P_n$ given by
\begin{subequations}
\label{eq:pnfirstfew}
\begin{align}
&P_0=\frac{\pi}{(-f_2)^{1/2}},\\
&P_1=\frac{3\pi f_3}{2(-f_2)^{3/2}},\\
&P_2=\frac{\pi\left[ 3(18+e^2)f_3^2-12(4+e^2)f_2f_4\right]}{16(-f_2)^{5/2}}.
\end{align}
\end{subequations}
Higher order $P_n$'s can also be computed but will not be shown here for their excessive length.

It is seen from Eq. \eqref{eq:deltaf2} that in general, as long as $p$ is large compared to the characteristic mass dimension of the spacetime, being it the mass, charge or other parameters, the above PS will converge. Moreover, since the process to obtain the coefficients $f_n$ of the function $F(u)$ is quite simple (although tedious) both logically and methodologically, we can attempt to solve the PS with quite high accuracy. Lastly, we also note that in the coefficients $f_n$ there are kinetic variables $(L,\,E)$, which in principle are related/redundant to parameters $(p,\,e)$ of the orbit. Therefore it will be better to replace them with the latter to show the dependence of the PS through a single pair of kinetic variables. This will be done in Subsec. \ref{subsec:compmeth}.

\subsubsection{The PNU method \label{subsubsec:pntwo}}

In Subsec. \ref{subsubsec:pnone}, to fit the true orbit, we modified the $\phi$ term in Eq. \eqref{eq:anzorder0} to a general function of $\psi(\phi)$ but kept the left-hand side unchanged. There however exist other PN methods which modify both $\phi$ and $u_0$.
We now present this method, which as stated previously, is inspired by the Poincare-Lighthill-Kuo method frequently used in asymptotic analysis \cite{plkmethod}.

In this method, we assume that the $u=1/r$ for the exact orbit can be expanded as a series of $1/p$
\begin{align}
u=\sum_{n=0}^{\infty}u_n p^{-(n+1)}, \label{eq:uinpexp}
\end{align}
with $u_n$ being coefficient functions of coordinate $\phi$. The $p$ at this stage is only a large radius, but otherwise ambiguous. Only after Eq. \eqref{eq:bd2dvalue} it will be identified as the semiletus rectum of the motion.
To find these $u_n$, we need to work out the equation that they satisfy first. Differentiating Eq. \eqref{eq:fuexpd} with respect to $\phi$, it becomes
\begin{align}
    \frac{\dd^2 u}{\dd \phi^2}-f_2 u=\frac{f_1}{2}+\sum_{n=3}\frac{nf_n}{2} u^{n-1} ,\label{eq: dU/dphi}
\end{align}
where we explicitly moved the term proportional to $u^1$ to the left-hand side. It is recognized that if the inhomogeneous terms on the right-hand side were absent in the above equation, it will permit a harmonic oscillatory solution and $-f_2$ will play the rule of some angular frequency squared, i.e., we can denote \begin{align}
-f_2\equiv 4\pi^2/\tau_0^2,~\text{i.e.,}~\tau_0=2\pi/\sqrt{-f_2}, \label{eq:tao0def}
\end{align}
where $\tau_0$ is the period of this harmonic oscillation. However, because of the presence of the inhomogenous terms, the true period for the periodic solution, denoted as $\tau$, will be modified from $\tau_0$. We assume that $\tau$ can be expanded as a series of $1/p$ too
\begin{align}
\tau=\sum_{n=0}^\infty\frac{\tau_{n}}{ p^n}.  \label{eq:omegainpexp}
\end{align}

Changing the variable from $\phi$ to $\psi$ by
\begin{align}
    \phi=\tau \psi, \label{eq:phipsi}
\end{align}
so that $\psi$ represent how many periods that $\phi$ has evolved, then multiplying Eq. \eqref{eq: dU/dphi} by $\tau^2$ and  substituting in expansions \eqref{eq:uinpexp} and \eqref{eq:omegainpexp}, we can collect this equation by order of $p$ and find a system of equations for $u_n$
\begin{subequations}
\begin{align}
{\frac{\dd^2u_0}{\dd\psi^2}}+u_0=&\frac{1}{-f_2}(\frac{pf_1}{2 })
       \label{equ:u0},\\
{{{\frac{\dd^2u_n}{\dd\psi^2}}+u_n}}=&\frac{pf_1}{8\pi^2}\sum_{l=0}^{n}\tau_{l}\tau_{n-l} +\frac{f_2}{4\pi^2}\sum_{k=0}^{n-1}\sum_{l=0}^{n-k}\tau_{l}\tau_{n-k-l}u_{k}\nn\\
+\sum_{k=3}^{n+2} \frac{kf_{k}}{8\pi^2} &\sum_{j=0}^{n+2-k}\prod_{i=1,l_i\in S_j}^{k-1}u_{l_{i}}\sum_{m=0}^{n+2-k-j}\tau_{m}\tau_{n+2-k-j-m}\nn\\
&~~(n=1,2,\cdots),
        \label{eq:un}
\end{align}
\end{subequations}
where each $l_i$ in the last line runs through all elements of $S_j$, a set of $k-1$ non-negative integers such that sum of its elements equals $j$. Note that in general $f_1$ is of order $1/p$ and therefore the $pf_1$ is included within the equation for $u_0$.

This is a system of second-order inhomogeneous linear equations. To solve them, we supplement the condition that at the apoapsis where $\psi=0=\phi$
\begin{align}
    u(0)=\frac1r=\frac{1-e}{p},~\frac{\dd u(0)}{\dd\psi}=0.
\label{eq:uncond1}\end{align}
Note that effectively in this {\it one} condition we used two variables $(p,\,e)$ rather than one, which means that there is one redundant real degree of freedom here. Strictly speaking, only when this condition is combined with the conditions \eqref{eq:bd2value} and \eqref{eq:bd2dvalue} at the other boundary of the radial motion (the periapsis), then we can interpret $p$ and $e$ as the motion's semilatus rectum and eccentricity.

For the $u_n$'s, condition \eqref{eq:uncond1} is equivalent to require
\begin{align}
&        \frac{\dd u_0(0)}{d\psi}=0,~ u_0(0)=1-e, \label{eq:bdu0}\\
&        \frac{\dd u_n(0)}{d\psi}=0,~ u_n(0)=0~~(n=1,2,\cdots). \label{eq:bdun}
    \end{align}
The solution to $u_0(\psi)$ is found to be
\begin{align}
u_0(\psi)=\left[ 1-e+\frac{p f_1}{2f_2}\right]\cos\psi -\frac{pf_1}{2f_2}.
\end{align}
Substituting this to Eq. \eqref{eq:un} for $n=1$, we will find that there are terms proportional to $(\cos\psi)^1$ in the inhomogeneous terms. Such terms will cause in the solution to $u_1(\psi)$ terms proportional to $\psi\cos\psi$. Similarly, if we continue the process, terms of even higher orders of $\psi$ such as $\psi^n\cos\psi~(n=1,2,\cdots)$ will appear in solutions to $u_n$. As $\psi$ evolves, these terms will cause the divergence of the corresponding $u_n(\psi\to\infty)$, which is physically not allowed for bounded orbits under consideration. Note that the existence of such terms is not affected by the boundary condition \eqref{eq:bdu0}. Instead, to eliminate these terms we have to directly force the coefficient of each such term to zero. From Eq. \eqref{eq:un} with $n=1$, this effectively establishes a linear equation for $\tau_1$ from which it is solved. Eventually, when we use the iteration to solve higher of $u_n$'s, the same reasoning will establish a linear system of $\tau_n~(n=1,2,\cdots)$ from which the entire perturbation series of $\tau$ can be solved.
Here for simplicity, we only list the first few $\tau_n$'s although the higher order ones are also easy to obtain
\begin{subequations}
\label{eq:tao12def}
    \begin{align}
\tau_1=&\frac{3\pi pf_1f_3}{2(-f_2)^{5/2}},\\
\tau_2=&\frac{3\pi ((1-e)p f_1+(1-e)^2f_2)(5f_3^2-4f_2f_4)}{8(-f_2)^{7/2}}\\
&+\frac{15\pi p^2 f_1^2(7f_3^2-4f_2f_4)}{32(-f_2)^{9/2}}.
\end{align}
\end{subequations}

With $\tau_1$ and solution to $u_0(\psi)$, we then can solve all $u_n(\psi)$ as well as $\tau_n~(n=1,2,\cdots)$ iteratively. Here we only list one more $u_n$ due to their excessive length
\begin{align}
u_1=&\frac{f_3}{16f_2^3}\left\{ -\left[ 3p^2f_1^2+4(1-e)pf_1f_2+4(1-e)^2f_2^2\right]\right.\nn\\
&+8\left[ p^2f_1^2+(1-e)pf_1f_2+(1-e)^2f_2^2\right]\cos\psi\nn\\
&\left.+\left[ pf_1+2(1-e)f_2\right]^2\cos (2\psi)\right\}.
\end{align}
We then can show that all these $u_n$ and therefore the entire exact orbit solution $u$
in Eq. \eqref{eq:uinpexp} is a periodic function of $\psi$ with period $2\pi$. Indeed, assuming that $u_i$ for $i$ from 1 to $n-1$ are linear combinations of $\cos(k\psi)~(k=0,1,\cdots,i+1)$
\begin{align}
    u_i(\psi)=\sum_{k=0}^{i+1} c_{ik} \cos(k\psi)~(i=0,\,1,\,\cdots,\,n-1),
\end{align}
where $c_{ik}$ are coefficients functions depending on $f_{j+1}$ and $\tau_{j-1}$ for $j$ from 1 $i$,
then it is a small exercise to show by induction that
$u_n(\psi)$ will also be a linear combination of the above form but with the upper index of the sum to be $n+1$.
After combining all such $u_n$'s, we have
\begin{align}
u(\psi)=\sum_{i=0}^\infty\sum_{k=0}^{i+1} c_{ik}\cos(k\psi) \equiv \sum_{n=0}^\infty  c_n\cos(n\psi),
\label{eq:pnusolu}
\end{align}
where we identified $\displaystyle c_n=\sum_{i=n-1}^\infty c_{in}~(n=0,\,1,\,\cdots)$ and we set $c_{-10}=0$.
Eq. \eqref{eq:pnusolu} clearly is a periodic function of $\psi$ with period $2\pi$.

With this knowledge, then from the correspondence \eqref{eq:phipsi} between $\phi$ with $\psi$,
we see that the motion is also a periodic function of $\phi$ with period $\tau$. Consequently, the PS using the PNU method is
\begin{align}
\delta_\pnu=&\tau-2\pi =\sum_{n=0}^\infty\frac{\tau_n}{p^n}-2\pi, \label{eq:deltaf3}
\end{align}
where $\tau_n$ were given in Eqs. \eqref{eq:tao0def} and \eqref{eq:tao12def}.

There are a few important comments that deserve mentioning. Although we use the same symbol $\psi$ in this subsection as in Subsec. \ref{subsubsec:pnone}, and that $\psi$ also goes from $0$ to $2\pi$ for one full cycle in the radial direction, this does not mean that the $\psi$ there equals the one here. Indeed, except for the boundary points, usually the $\psi$ in the last subsection will not equal the $\psi$ here which is a linear change of $\phi$. This also means that the true exact orbit in this subsection does not equal to
\begin{align}
    u\neq \frac{1}{p}+\frac{e \cos(\phi/\tau)}{p}
\end{align}
for general $\phi$, except perhaps the boundaries. The second comment is that similar to the PS with PN$\Phi$ method given in Eq. \eqref{eq:deltaf2}, the result \eqref{eq:deltaf3} also contains the redundant kinetic variables $(L,\,E)$, which will be replaced later using the method introduced in Subsec. \ref{subsec:compmeth}.

\subsection{Equivalence of the methods
\label{subsec:compmeth}}

As mentioned in Subsecs. \ref{subsec:grse} and \ref{subsec:pn}, the SE method result is most accurate in the small eccentricity $e\to0$ limit while the latter two PN method works best in the large radius $p\to\infty$ limit.

In the PN results \eqref{eq:deltaf2} and \eqref{eq:deltaf3}, the expansion coefficients $f_n$ of the function $F(u)$ depend on, besides the spacetime parameters, the orbit kinetic parameters $L$ and $E$ too. On the other hand, the other terms in these formulas contain the semilatus rectum $p$ and eccentricity $e$, which are essentially equivalent to $L,~E$. In order to show the dependence of the PS on the orbit kinematics through only one pair of variables, this redundancy has to be eliminated. Since many previous computations and experiments have used the PN variables $(p,\,e)$, we will transform the $(L,\,E)$ in our results to $(p,\,e)$ too.

To do this, we will use the boundary conditions that at $r_1$ and $r_2$
\begin{subequations}
    \begin{align}
&u(r_1)\equiv\frac{1}{r_1}=\frac{1+e}{p},\\
&u(r_2)\equiv\frac{1}{r_2}=\frac{1-e}{p}. \label{eq:bd2value}
\end{align}
\end{subequations}
Moreover, since they are the turning points in the radial direction, from Eq. \eqref{eq:fuexpd} we should also have at these boundaries
\begin{subequations}
\label{eq:bddvalue}
\begin{align}
F\left(\frac{1}{r_1}\right)=F\lb \frac{1+e}{p}\rb =0,\\
F\left(\frac{1}{r_2}\right)=F\lb \frac{1-e}{p}\rb =0.\label{eq:bd2dvalue}
\end{align}
\end{subequations}
The conditions \eqref{eq:bd2value} and \eqref{eq:bd2dvalue}  are used for the first time for the PNU results.
Forcing these conditions onto the solution in this method effectively removes the redundancy mentioned after Eq. \eqref{eq:uncond1}.
In other words, we can now interpret the $p$ and $e$ in the PNU method as the semilatus rectum and eccentricity respectively.

Substituting the definition of $F(u)$ in Eq. \eqref{eq:fudef} into Eqs. \eqref{eq:bddvalue}, we obtain two equations linking  $(L,\, E)$ and $(p,\,e)$, whose solution in principle should provide the transformation we desired.
However, for general metric functions and therefore $F(u)$, it is not possible to obtain the explicit solutions to $(L,\,E)$ in terms of $(p,\,e)$. Fortunately, since all our computations are perturbative, a series solution suffices. A simple PN order estimation shows that we can assume that $(L,\,E)$ take the following series form when $p$ is large
\begin{align}
    E=\sum_{n=0}^\infty \frac{e_n}{p^{n/2}},~L=\sum_{n=-1}^\infty \frac{l_n}{p^{n/2}}.
    \label{eq:elexpinpe}
\end{align}
Substituting these into conditions \eqref{eq:bddvalue} and using the asymptotic expansions \eqref{eq:metricexp}, the coefficients $e_n,\,l_n$ can be found using the undetermined coefficient method. The results for the first few orders are
\begin{subequations}
\label{eq:elexpcoeff}
    \begin{align}
e_0=&1,\\
e_2=&\frac{a_1(1-e^2)}{4},\\
e_4=&\frac{a_1(3a_1-4c_1)(1-e)^2}{32},\\
e_5=&\frac{\sqrt{-a_1}b_1(1-e^2)^2}{4\sqrt{2}},\\
\cdots&~\cdots,\nn\\
l_{-1}=&\sqrt{\frac{-a_1}{2}},\\
l_1=&-\frac{4a_2-a_1(a_1-c_1)(3+e^2)}{4\sqrt{-2a_1}},\\
l_2=&-\frac{b_1(3+e^2)}{4},\\
\cdots&~\cdots.\nn
    \end{align}
\end{subequations}
Here it is found that $e_1=e_3=l_0=0$ in general, and higher order terms can be obtained easily.
Substituting the $L,\,E$ with these coefficients into the $f_n$'s in Eq. \eqref{eq:fexpcoeff}, and then eventually into the PS \eqref{eq:deltaf2} and \eqref{eq:deltaf3}, the PS in these two PN methods can all be expressed in terms of parameter $p,\, e$.
It is found that the two PN methods yield exactly the same series of $p$
\begin{align}
\delta_{\text{PN}}=&\frac{\pi[-2a_1^2+(c_1+d_1)a_1+2a_2]}{a_1p} +\frac{2\sqrt{2}\pi b_1}{\sqrt{-a_1}p^{\frac{3}{2}}}\nn\\
&+\frac{\pi}{8a_1^2p^2}\{40a_1^4+8[(e^2-4)c_1-d_1]a_1^3-8a_2^2\nn\\
&-[(2+5e^2)c_1^2+2(-2+e^2)d_1c_1-4(8+e^2)c_2\nn\\
&+(2+e^2)(d_1^2-4d_2)+80a_2]a_1^2\nn\\
&+8[6a_3+a_2((4-e^2)c_1+d_1)]a_1\}\nn\\
&+\frac{\pi}{\sqrt{2}(-a_1)^{\frac{3}{2}}p^{\frac{5}{2}}}\{16a_1^2b_1+2(-3+e^2)a_2b_1\nn\\
&-a_1[2(5+e^2)b_2+b_1c_1(5-3e^2)+2b_1d_1]\}\nn\\
&+\calco(p)^{-3},
\label{eq:deltaf4}
\end{align}
where $a_i,\,b_i,\,c_i,\,d_i$ are coefficients of the asymptotic expansion \eqref{eq:metricexp}. Of course, this is no surprise because after all, both PS found using respectively the PN$\Phi$ and PNU methods are essentially the same kind of PN expansions in terms of large $p$.
Therefore from now on, we will not distinguish the results \eqref{eq:deltaf2} and \eqref{eq:deltaf3} but directly use the above Eq. \eqref{eq:deltaf4} for the PN PS.
Eq. \eqref{eq:deltaf4} is one of the key results of this work. Its higher-order terms have also been worked out but are too lengthy to present here. This result provides the PS of test particles in the equatorial plane for general SAS spacetime. All dependence of the PS on spacetime parameters can be read off from this formula. For example, since usually $-a_1=2M$ and $b_1=-4\hat{a}M$ where $M$ and $\hat{a}$ are the spacetime mass and spin, from the $p^{-3/2}$ term we recognize that it is the spin $\hat{a}$ and {\it only} $\hat{a}$ ($M$ is treated as a base scale) that will contribute to the frame-dragging effect of the PS in all SAS spacetimes.

If we want to compare the PN results with that obtained using the SE method, besides converting $(L,\,E)$ to $(p,\,e)$, we also need to transform the $r_c$ appearing in $y_n$ and $\eta$ in Eq. \eqref{eq:deltaf1} to these two variables. Although converting $r_c$ to a series of $p$ can be done to infinite accuracy, when we substitute the series for the $r_c$ in $y_n$ and $\eta$ in Eq. \eqref{eq:deltaf1} and truncate the result $\delta_{\text{SE}}$ for finite order of $\eta$, the result is only valid in the simultaneous limits $p\to\infty$ and $e\to 0$. And this is also the part of the parameter space for which we can compare the SE method result with those of the PN method.

When $r_c$ or $p$ is large, it is trivial to expect that to the leading order $r_c\propto p$. Therefore we can assume that $r_c$ can be expanded as a series of $p$
\begin{align}
    r_c=\sum_{n=-2}^\infty\frac{w_n(e)}{p^{n/2}}, \label{eq:rcexp}
\end{align}
where $w_n(e)$ are functions of the eccentricity. Substituting this, and the series \eqref{eq:elexpinpe} for $(L,\, E)$ as well as \eqref{eq:metricexp} for the metric functions into
the defining Eq. \eqref{eq:dlreq} for $r_c$, we can solve perturbatively all coefficient functions $w_n$ for the $r_c$ series \eqref{eq:rcexp}. The result to the first few orders are
\begin{subequations}
\begin{align}
w_{-2}=&\frac{1}{1-e^2},\\
w_{-1}=&0,\\
w_{0}=&0,\\
w_1=&\frac{(1-\sqrt{1-e^2})b_1}{\sqrt{-2a_1}},\\
w_2=&\frac{e^2 \left(a_1^3-a_1^2
   c_1+a_1 (c_2-2
   a_2)+a_2
   c_1+a_3\right)}{a_1},\\
w_3=&\frac{(4a_2b_1-12a_1^2b_1+8a_1b_2+5a_1b_1c_1)(1-(1-e^2)^{\frac{3}{2}})}{4\sqrt{2}(-a_1)^{\frac{3}{2}}}\nn\\
&+\frac{e^2 a_1b_1c_1}{4\sqrt{2}(-a_1)^{\frac{3}{2}}}.
\end{align}
\end{subequations}

With $(L,\,E)$ and $r_c$ expressed as series of $p$ in Eqs. \eqref{eq:elexpinpe} and \eqref{eq:rcexp}, we can now attempt to express the SE method PS \eqref{eq:deltaf1} also into series of $p$. To do this, we express $\eta$ and $y_n$ as series of $p$ first.
Substituting Eqs. \eqref{eq:metricexp}, \eqref{eq:elexpinpe} and \eqref{eq:rcexp} into the definition \eqref{eq:etadef} of $\eta$, we find that
\begin{align}
\eta=&\frac{1}{\sqrt{1-e^2}}-1-\frac{(c_1-a_1+a_2/a_1)e^2}{\sqrt{1-e^2}}\frac{1}{p}\nn\\
&+\frac{b_1[(1-e^2)^{\frac{3}{2}}-(1+e^2)]}{\sqrt{-2(1-e^2)a_1}}\frac{1}{p^{\frac{3}{2}}}.
\label{eq:etaexp}
\end{align}
It is seen that in general, if $e$ is not small, the $\eta$ will have a nonzero zero-order term in its large $p$ expansion. Only in the $e\to0$ will $\eta$ approaches zero.
For the coefficients $y_n$, their $p$-series expansions are more involved. Although we can still find their series with the help of a computer algebraic system, here we will not list explicitly their general forms but only state their relevant features:
$y_n$ for general $n$ are found to have nonzero $p^0$-order coefficient too.
Therefore, after substituting these $y_n$ and the above series \eqref{eq:etaexp} into result \eqref{eq:deltaf1}, the
total PS will not necessarily converge if no other limit is taken. Again, as we mentioned before, this is consistent with the fact that the SE method is only valid when the eccentricity is small. Further carrying out the small $e$ expansion, the result for the PS becomes
\begin{align}
    \delta_{\text{SE}} =\sum_{n=2}^\infty\sum_{m=0}^{\left[\frac{n-1+(-1)^n}{4}\right]} z_{nm}\frac{e^{2m}}{p^\frac{n}{2}},
\end{align}
where the coefficients are
\begin{subequations}
    \begin{align}
        z_{20}=&\frac{\pi(-2a_1^2+2a_2+a_1c_1+a_1d_1)}{a_1},\\
        z_{30}=&\frac{2\sqrt{2}\pi b_1}{\sqrt{-a_1}},\\
        z_{40}=&\frac{\pi}{4a_1^2}\{20a_1^4-4a_1^3(4c_1+d_1)+4a_1[6a_3+a_2(4c_1+d_1)]\nn\\
        &-a_1^2(40a_2+c_1^2-16c_2-2c_1d_1+d_1^2-4d_2-4a_2^2)\}\nn\\
        z_{41}=& \frac{\pi[8a_1^2c_1-a_1(5c_1^2-4c_2+2c_1d_1+d1^2-4d_2)-8a_2c_1]}{8a_1}.
    \end{align}
    \end{subequations}
One can easily verify that this is indeed the same as result \eqref{eq:deltaf4} after taking the small $e$ expansion there.
This consistency also confirms the validity of the SE method in general.

In physical systems with PS data and to which the PS formulas above are applicable (e.g., satisfying the test particle assumption), all $p$'s are much larger than $M$ and therefore the large $p$ limit works. Therefore in Sec. \ref{sec:appinastro}, we will use extensively the large $p$-limit formula \eqref{eq:deltaf4} of the PS. However, we remind the reader that this does not mean that results there are exclusively obtained using the PN approach. As shown in this section, when the SE formula \eqref{eq:deltaf1} is further expanded to large $p$, it also reproduces the PN formula \eqref{eq:deltaf4}.

\section{Applications in particular spacetimes \label{sec:appinspacetimes}}

In this section, we directly apply the two methods described above to some known spacetimes to find the PS in them. We will compare with known results in the literature and study the effect of various parameters on these shifts.
%For each of the following spacetimes, we will present the shifts in the order of the SE, PN$\Phi$ and PNU methods. The corresponding results are in Eqs. \eqref{eq:deltaf1}, \eqref{eq:deltaf2} and \eqref{eq:deltaf3} respectively.

\subsection{PS in KN, Kerr, RN and Schwarzschild spacetimes\label{subsec:kn}}

The KN line element in the Boyer-Linquidist coordinate is given by Eq. \eqref{eq:metric} with metric functions
\begin{align}
&A(r)=1-\frac{2 M r-Q^2}{\Sigma},~    B(r)=-\frac{2 \hat{a} \sin ^2\theta \left(2 M r-Q^2\right)}{\Sigma},\nn\\
&C(r)=  \left[{\frac{(r^2+\hat{a}^2)^2-\Delta \hat{a}^2\sin^2\theta}{\Sigma}}\right]\sin^2\theta,~
D(r)=\frac{\Sigma }{\Delta },    \label{eq:knmetric}
\end{align}
where $M,\, \hat{a},\,Q$ are the mass, spin angular momentum per unit mass of the BH and the BH charge, and
\begin{align}
\Sigma =r^2+\hat{a}^2\cos^2\theta,~
\Delta=r^2-2Mr+\hat{a}^2+Q^2. \label{eq:sigmadeltakndef}
\end{align}

For the SE method, to write the PS using \eqref{eq:deltaf1}, what is needed is the expansion of the metrics around the extreme point $r_c$ of $l(r)$. Substituting metric \eqref{eq:knmetric} and then $\theta=\pi/2$ for the equatorial plane into Eq. \eqref{eq:dlreq}, the equation fixing the $r_c$ for the KN spacetime is found to be an eight-order polynomial of $r_c$ from which we can not yield an explicit analytical solution. For the Kerr case with $Q=0$, this equation will reduce to a six-order polynomial whose explicit solution still can not be given here. Only when we further set $\hat{a}=0$ for Schwarzschild spacetime, this $r_c$ can be solved to find
\begin{align}
r_{c,\text{S}}=  \frac{\left(4-3 E^2+\sqrt{9 E^2-8} E\right) M}{2 (1-E^2)}
.\label{eq:schrcexp}
\end{align}
Therefore for the full KN case, we will only assume that the radius $r_c$ is solved numerically once other spacetime and particle parameters are given.

At this $r_c$, the constant $\eta$ defined in Eq. \eqref{eq:etadef} is found as
\begin{align}
    \eta=&\left[ r \sqrt{\left(\hat{a}^2-2 M r+Q^2+r^2\right) \left((E^2-1) r^2+2 Mr+Q^2\right)}\right.\nn\\
    &+  E\hat{a}( Q^2 -2 M r)\Big] \frac{1}{L\left[ r (r-2 M)+Q^2\right]}\bigg|_{r=r_c}-1,
    \label{eq:etakn}
\end{align}
where the definition of $l(r)$ in Eq. \eqref{eq:lrdef} for the KN case was used. On the other hand, at the same $r_c$,
the expansions of the metric functions \eqref{eq:knmetric} are found as
\begin{subequations}
    \begin{align}
        A(r)=&1+\frac{Q^2-2 M r_c}{r_c^2}+\frac{2  \left(Q^2-M r_c\right)}{r_c}s+Q^2 s^2,\nn\\
        B(r)=&\frac{2 \hat{a} \left(Q^2-2 M r_c\right)}{r_c^2}+\frac{4 \hat{a} \left(Q^2-M r_c\right)}{r_c}s+2 \hat{a} Q^2 s^2\nn\\
        C(r)=&\frac{\hat{a}^2 \left(2 M r_c-Q^2+r_c^2\right)+r_c^4}{r_c^2}+\frac{2[ \hat{a}^2 \left(M r_c-Q^2\right)-r_c^4]}{r_c}s\nn\\
        &+(3r_c^4- \hat{a}^2 Q^2) s^2 +\sum_{i=3}^{\infty}(-1)^i(i+1) r_c^{i-2} s^{i},\nn\\
        D(r)=&\frac{r_c^2}{\hat{a}^2-2 M r_c+r_c^2+Q^2}-\frac{2r_c^3\left( \hat{a}^2 -M r_c+Q^2 \right)}{\left(\hat{a}^2-2 M r_c+r_c^2+Q^2\right)^2}s\nn\\
        &+\frac{\left[\left(M^2-\hat{a}^2-Q^2\right)r_c^2  +3(\hat{a}^2+Q^2-Mr_c)^2\right]r_c^4}{\left(M^2-2 M r_c+r_c^2+Q^2\right)^3}s^2\nn\\
&        +\calco(s)^3, \nn
    \end{align}
\end{subequations}
where $\displaystyle s=\frac1r-\frac{1}{r_c}$ and it is seen that the series for $A(r)$ and $B(r)$ are indeed finite sums.
From these expansion we can read off the coefficients $a_i,\,b_i,\,c_i,\,d_i$ in Eq. \eqref{eq:abcdexpatrc}.

\begin{figure}[htp!]
\centering
\includegraphics[width=0.22\textwidth]
{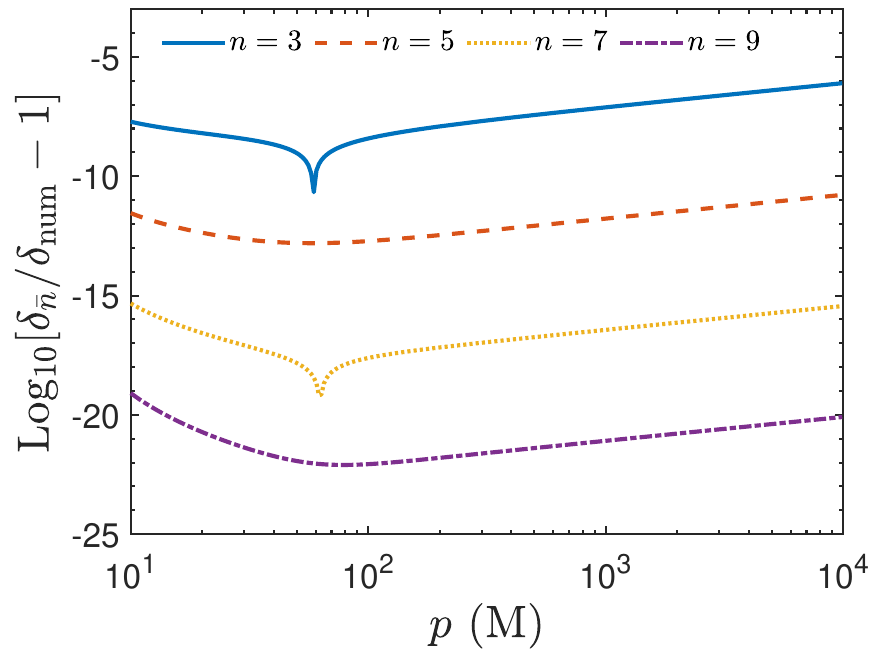}\includegraphics[width=0.22\textwidth]
{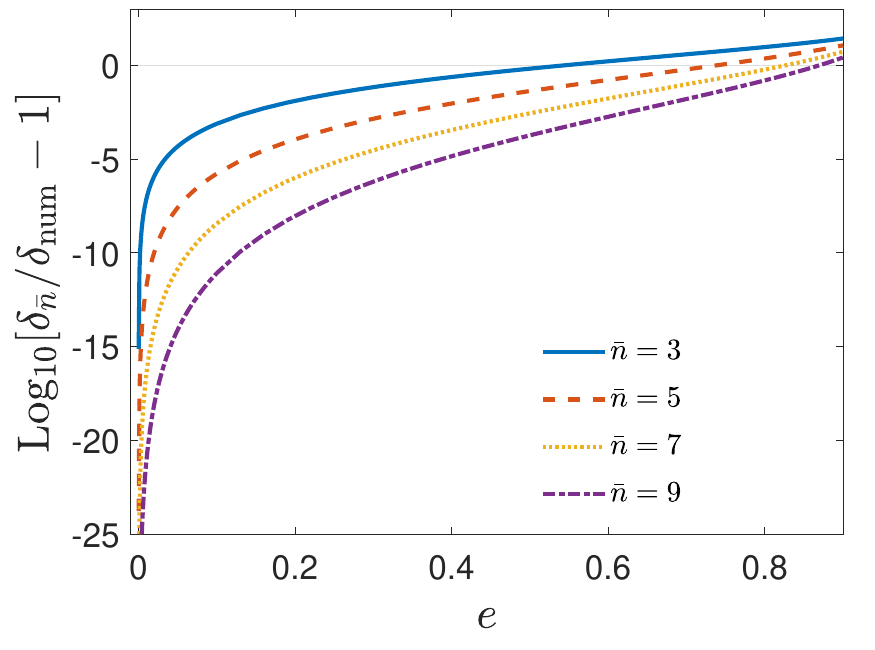}\\
(a)\hspace{3cm}
(b)\\
\includegraphics[width=0.22\textwidth]
{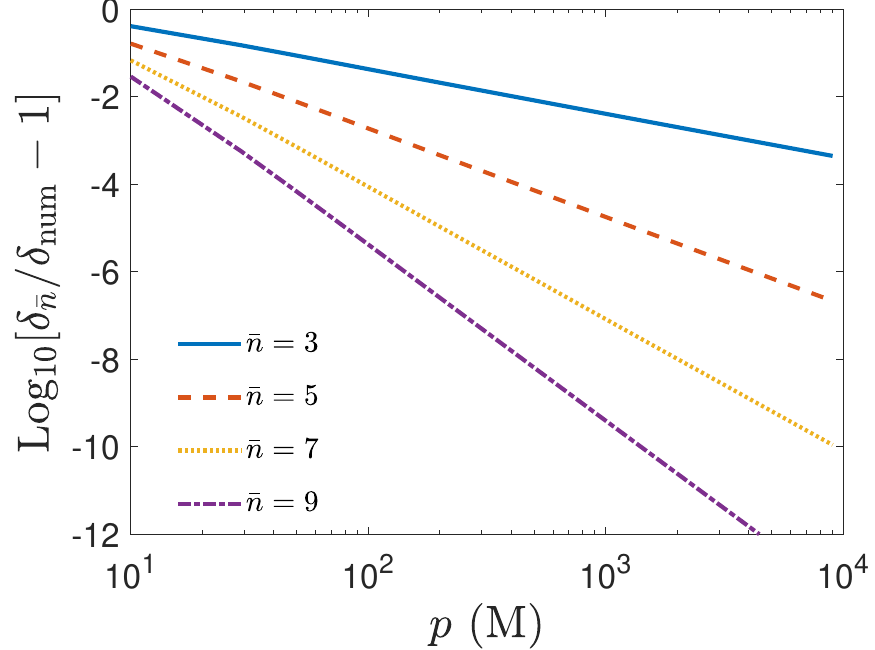}
\includegraphics[width=0.22\textwidth]
{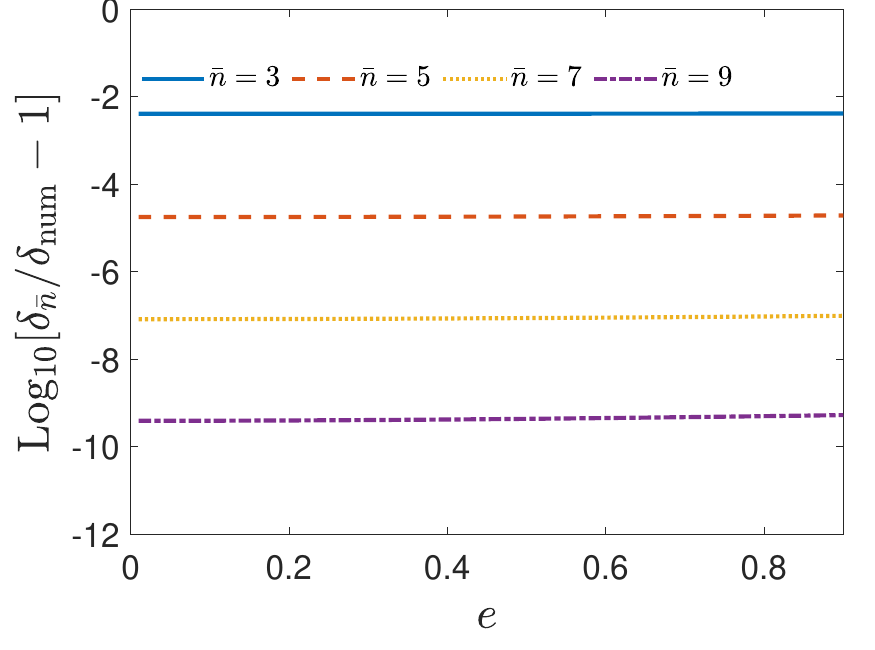}\\
(c)\hspace{3cm}
(d)\\
%\includegraphics[width=0.22\textwidth]
%{cop.pdf}
%\includegraphics[width=0.22\textwidth]
%{coe.pdf}
%\\
\includegraphics[width=0.22\textwidth]
{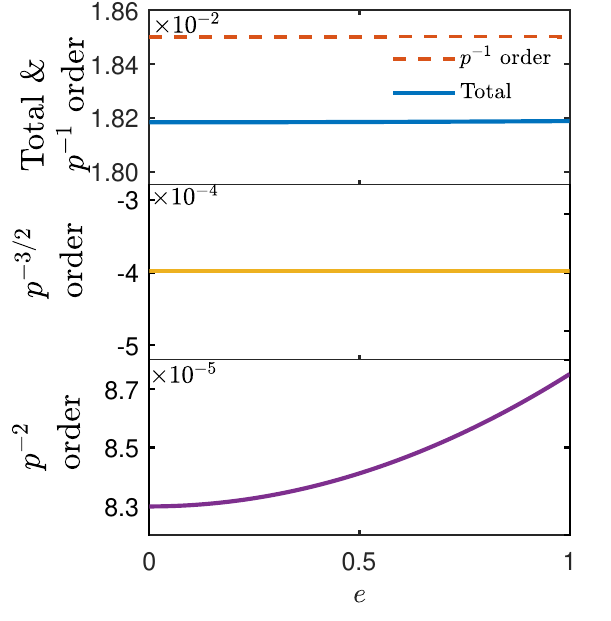}
\includegraphics[width=0.22\textwidth]
{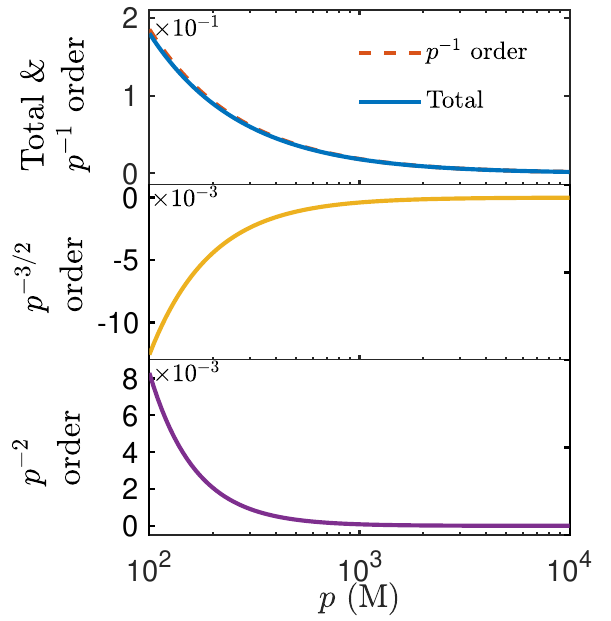}\\
(e)\hspace{3cm}
(f)\\
%\includegraphics[width=0.22\textwidth]
%{coa.pdf}
%\includegraphics[width=0.22\textwidth]
%{coQ.pdf}
%\\
\includegraphics[width=0.22\textwidth]
{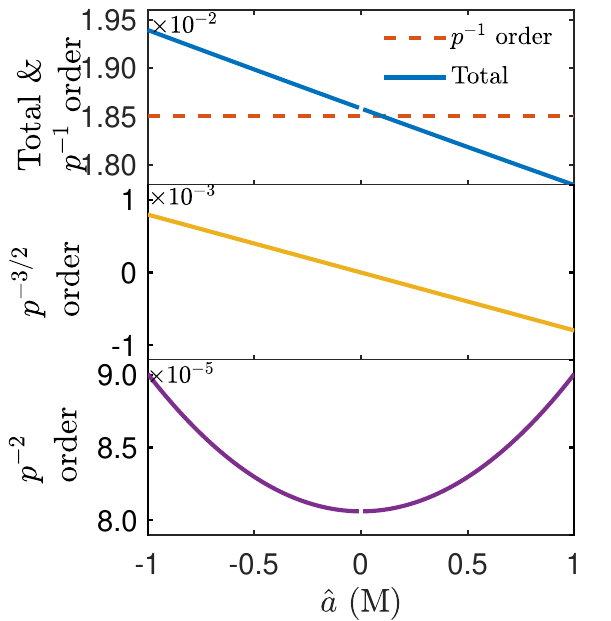}
\includegraphics[width=0.22\textwidth]
{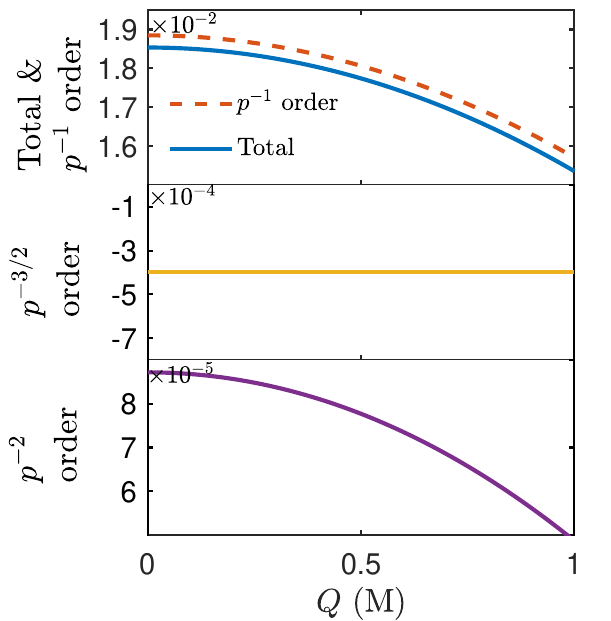}\\
(g)\hspace{3cm}
(h)
\caption{PS in KN, Kerr, RN and Schwarzschild spacetimes. Comparison of the numerical result with the SE method result ((a) and (b)), and with the PN result ((c) and (d)). The dependence of the PS on $e$ (e), $p$ (f), $\hat{a}$ (g) and $Q$ (h) (top panel: total and leading order PS; middle panel: sub-leading PS; bottom panel: third order PS). The default parameters are $p=1000M,~e=1/100,~Q=M/3,~\hat{a}=M/2,~M=1$ except the one being varied. }
   \label{fig:kn}
\end{figure}

Substituting these coefficients into Eq. \eqref{eq:yexp}, we can find the corresponding $y_n$ in the KN spacetime. Since in general all expansion coefficients are present, the expressions for the $y_n$ are not shortened by substituting expressions of the coefficients $a_i,\,b_i,\,c_i,\,d_i$. Therefore we will not write $y_n$ out here.
Substituting these $y_n$'s, as well as Eq. \eqref{eq:etadef} into Eq. \eqref{eq:deltaf1}, then the PS in the KN spacetime using the SE method is obtained. We will show the properties of this PS in Fig. \ref{fig:kn} after we obtain the PN PS next.

To find the PS in the PN methods, we first expand the metric functions according to Eq. \eqref{eq:metricexp} for the KN spacetime
\begin{subequations}
\label{eq:knasyexp}
    \begin{align}
        A(r)=&1-\frac{2 M}{r}+\frac{Q^2}{r^2},\\
        B(r)=&-\frac{4 \hat{a} M}{r}+\frac{2 \hat{a} Q^2}{r^2},\\
        C(r)=&r^2+\hat{a}^2+\frac{2 \hat{a}^2 M}{r}-\frac{\hat{a}^2 Q^2}{r^2},\\
        D(r)=&1+\frac{2 M}{r}+\frac{-\hat{a}^2+4 M^2-Q^2}{r^2}+\calco(r)^{-3},
    \end{align}
\end{subequations}
from which the asymptotic coefficients $a_n,\,b_n,\,c_n,\,d_n$ can be read off.
%Further substituting into Eqs. \eqref{eq:fexpcoeff}, we find
%\begin{subequations}
%\label{eq:fnkn}
%\begin{align}
%f_{0,\text{KN}}=&\frac{E^2-1}{L^2},\\
%f_{1,\text{KN}}=&\frac{2M}{L^2}-\frac{4\hat{a}ME(E^2-1)}{L^3},\\
%f_{2,\text{KN}}=&-1+\frac{3\hat{a}^2(E^2-1)-Q^2}{L^2}+\frac{12\hat{a}^2E^2(E^2-1)M^2}{L^4}\nn\\
%&-\frac{2\hat{a}E[4E^2M^2-Q^2(E^2-1)]}{L^3}.    \end{align}
%\end{subequations}
%To facilitate later comparison between different methods, for the KN spacetime the $(L,\,E)$ here can be transformed to $(p,\,e)$ using Eq. \eqref{eq:elexpcoeff} with the help of the asymptotic expansions \eqref{eq:knasyexp}. The corresponding results for the first few orders are
%\begin{subequations}\label{eq:elfirstfewkn}
%    \begin{align}
%        e_0=&1,\\
%        e_2=&\frac{1}{2}(e^2-1) M,\\
%        e_4=&\frac{3}{8}(e^2-1)^2 M^2,\\
%        e_5=&-\hat{a} \left(e^2-1\right)^2 M^{3/2},\\
%        e_6=&\frac{M(e^2-1)^2}{16} \left[8 \hat{a}^2+M^2(5 e^2 +27)-16 Q^2\right]   ,\\
%        l_{-1}=&\sqrt{M},\\
%        l_1=&\frac{e^2 M^2+3 M^2-Q^2}{2 \sqrt{M}},\\
%        l_2=&-\hat{a} \left(e^2+3\right) M,\\
%        l_3=&\frac{3M^{5/2}}{8} \left( e^2+3\right)^2 +\sqrt{M} \left[ \hat{a}^2\left(e^2+1\right)\right.\nn\\
%        &\left.-\left(5 e^2+7\right) Q^2/4\right]-\frac{Q^4}{8 M^{3/2}},\\
%        l_4=&\frac{-\hat{a} M^2 }{2} \left(e^2+3\right) \left(3 e^2+5\right)+2\hat{a} Q^2 \left(e^2+1\right).
%    \end{align}
%\end{subequations}
%And we remind that $e_1,\, e_3,\, l_0$ are zero in general.
After substituting the coefficients in Eq. \eqref{eq:knasyexp} into  Eq. \eqref{eq:deltaf4}, we immediately obtain the PN result to the first few orders in KN spacetime
\begin{align}
\delta_{\text{PN,KN}}=&\frac{\pi M}{p}\left[6-\left(\frac{Q}{M}\right)^2\right]- 8\pi\left(\frac{M}{p}\right)^{3/2}\frac{ \hat{a}}{M} \nn\\
&+\frac{\pi}{4 } \left(\frac{M}{p}\right)^{2}\left[6(18+e^2) +12\left(\frac{\hat{a}}{M}\right)^2\right.\nn\\
&\left.-2(24+e^2)\left(\frac{Q}{M}\right)^2-\left(\frac{Q}{M}\right)^4\right]\nn\\
        &- 8\pi\left(\frac{M}{p}\right)^{5/2}\left[ 9\frac{\hat{a}}{M} -2\left(\frac{Q}{M}\right)^2\right]\nn\\
        &+\frac{\pi}{8}\left(\frac{M}{p}\right)^3\left[-180\left(6+e^2\right)\right.\nn\\
&+24\left(3e^2-25\right)\left(\frac{\hat{a}}{M}\right)^2-4\left(e^2-11\right)\left(\frac{\hat{a}Q}{M^2}\right)^2\nn\\
&+6\left(126+19e^2\right)\left(\frac{Q}{M}\right)^2+2\left(e^2-31\right)\left(\frac{Q}{M}\right)^4\nn\\
        &\left.+\left(\frac{Q}{M}\right)^6\right]+\mathcal{O}(p^{-7/2}).
        \label{eq:pnreskn}
\end{align}
Higher orders are not presented explicitly because of their length.

We note that this result when truncated to low orders agrees with Eq. (16) of Ref. \cite{Jiang2014} (order $p^{-3/2}$), Eq. (23) of Ref.\cite{gong2008} and Eq. (35) of Ref. \cite{Heydari-Fard:2019kno} (order $p^{-2}$) which only computed for zero eccentricity. When setting $Q=0$ in Eq. \eqref{eq:pnreskn}, it reduces to the PS in Kerr spacetime. And the Kerr result when truncated to order $p^{-3/2}$ agrees with Eq. (4.11) of Ref. \cite{Brill1997Light} and to order $p^{-2}$ and $p^{-5/2}$ with Eq. (40) of Ref. \cite{Colistete:2002ka} and Eq. (21) of Ref. \cite{LeTiec:2013uey} (both only computed the case of zero eccentricity and the results are not very explicit) respectively.
If we set $\hat{a}=0$ in Eq. \eqref{eq:pnreskn}, the PS in RN spacetime for neutral particles is obtained. Such obtained RN result to the lowest $p^{-1}$ order agrees with corresponding results in Refs.
\cite{Gong:2009zzb,A2011The,Hu:2013eya,Shchigolev:2016vpz,Uniyal:2017yll,Mak:2018hcn,Bakry:2021smq,Zhang:2022zox,Mandal2022}.
Finally, if both $Q=\hat{a}=0$ are set, Eq. \eqref{eq:pnreskn} yields the Schwarzschild result, whose $p^{-1}$ order (or at most order $p^{-2}$) was studied in Ref. \cite{Colistete:2002ka,2006Precession,Shchigolev:2015sgg,Shchigolev:2016vpz,Mak:2018hcn,Walter2018,Hall:2022ekh,Tucker:2019CQG}.
For future reference, we provide the high orders of the PS in the Kerr, RN and Schwarzschild spacetimes in the Appendix. \ref{sec:apphigh}.

Now to verify the correctness of our method, and to study the effect of the spacetime parameters,
%especially the spin $a$ on the PS,
we plot in Fig. \ref{fig:kn} the dependence of the $\delta_{\text{SE,KN}}$ and  $\delta_{\text{PN,KN}}$ on the kinetic variables $(p,\,e)$ as well as $\hat{a}$ and $Q$.
In Fig. \ref{fig:kn} (a) and (b), we plot as functions of $p$ and $e$ the difference between the PS obtained numerically by directly integrating the definition Eq. \eqref{eq:deltadef} and that using the SE method, i.e. Eqs. \eqref{eq:deltaf1}, but truncated to order $\bar{n}$
\begin{equation}
    \delta_{\text{SE},\bar n}=2\sum_{n=0}^{\left[\frac{\bar{n}}{2} \right]}y_nI_n-2\pi.
\end{equation}
The $(L,\,E,\, r_c)$ here are replaced by $(p,\,e)$ numerically.
Similarly, in Fig. \ref{fig:kn} (c) and (d), the difference between the numerical result and the PN result
\begin{equation}
    \delta_{\text{PN},\bar n}=2\sum_{n=0}^{\bar n}\frac{P_n}{p^{n/2}}-2\pi,
\end{equation}
to order $\bar{n}$ is plotted. As long as the numerical integration is carried out to high enough accuracy, its result can be viewed as the true PS against which our perturbative results can be compared. Note that in this figure, as well as Fig. \ref{fig:ks} and Fig. \ref{fig:ktn}, some range of the parameters does not correspond to BH spacetime anymore. However these plots, as well as the formulas in this work are still valid because the PS are not qualitatively sensitive to the nature of the inner spacetime.

It is seen that as the truncation order increases, the results using both the SE and PN methods converge to the numerical PS exponentially. Moreover, for the SE method result which essentially is a small $e$ expansion, from plots (a) and (b) we see that the smaller the eccentricity, the better it approximates the true PS. For the parameters chosen in these plots, when $e\gtrsim 0.85 $, the highest order approximation ($\bar{n}=9$) will not be reliable. While as $p$ increases, the performance of the SE result roughly maintains constant for fixed $\bar{n}$. In contrast, the PN method result works much better when $p$ is large due to the fact that it is a large $p$ expansion. For the chosen parameters, when $p\gtrsim 1000M$, the highest order approximation ($\bar{n}=9$) can maintain an accuracy of $10^{-10}$, enough even for all current astrophysical applications. Moreover, the dependence of the PN result's performance on $e$ is so small that it can not be recognized in Fig. \ref{fig:kn} (d) by eye inspection.

Since plot (a)-(d) verified the high accuracy of the perturbative PS, in Fig. \ref{fig:kn} (e)-(h) we plot the dependence of the total PS found using our perturbative method to order 9, as well as contributions from the leading three orders of the PN result, on kinetic parameters $(p,\,e)$ and spacetime parameters $\hat{a}$ and $Q$.
It is seen for all four parameters, the leading order dominates the size of the PS. For the effect of $e$, the first non-trivial effect to the PS comes from the $p^2$ order. While for the parameter $p$, its effect manifests from the leading order and this order dominates higher orders.

The effect of parameters $\hat{a}$ and $Q$ on the PS are more interesting because they are purely GR effects. It is seen from the top panel of Fig. \ref{fig:kn} (g) that
the leading order (order $p^{-1}$) contribution is independent of $\hat{a}$. This is indeed the famous Schwarzschild-like (perhaps a more precise name when the charge is nonzero, is the RN-like) PS, which Einstein used to explain the extra PS of Mercury around the Sun. The non-trivial dependence on $\hat{a}$ starts from the next-to-leading order (order $p^{-3/2}$), which is also referred to as the frame-dragging (FD) (or the Einstein-Lense-Thirring, or gravitomagnetic) term. The middle panel of Fig. \ref{fig:kn} (g) shows that the PS of a corotating (or counter-rotating) orbit will be decreased (or increased) by the spacetime spin. This effect is in alignment with the effect of spacetime spin on the deflection angle of signals traveling in the equatorial plane \cite{Huang:2020trl,Jia:2020xbc}.
There, the effect of $\hat{a}$ also appears from the second non-trivial order. The third order (order $p^2$) depends only on $\hat{a}^2$ and therefore will increase the PS when it deviates from zero, regardless of the signal rotation direction.

The effect of the spacetime charge is a less studied parameter because of the general expectation by astronomers that most celestial objects are neutral. Fig. \ref{fig:kn} (h) top panel shows that unlike the FD effect, the spacetime charge decreases the PS at the very leading order. Although this trend is consistent with the effect of $Q$ on the deflection angle of signals in RN spacetime \cite{Pang:2018jpm}, the order $Q$ manifests itself is one order lower than in the deflection angle. There, $Q$ only influences the deflection from the second non-trivial order \cite{Jia:2020xbc}. For the parameters chosen when drawing these plots, and if $|Q|\leq 1$ for a BH spacetime, the charge effect to the PS can reach a large fraction of the total PS. At the $p^{-3/2}$ order, the charge has no contribution while at the
$p^{-2}$ order, it will decrease the PS by a small amount.

\begin{figure}[htp!]
\centering
\includegraphics[width=0.54\textwidth]
{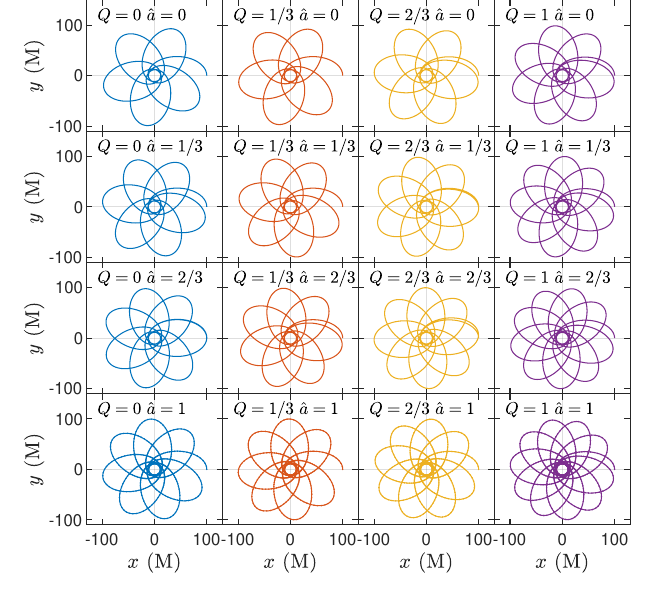}
%\\
%\includegraphics[width=0.45\textwidth]{orbit.pdf}
\caption{Counter-clockwise rotating orbits in KN spacetime with various values of $\hat{a}$ and $Q$. The initial point is on the positive $x$-axis. Other parameters are $p=20M,~e=4/5,~M=1$. }
\label{fig:knorbits}
\end{figure}

All of the above-mentioned dependence on various parameters can be understood by analyzing the total PS given in  Eq. \eqref{eq:pnreskn}. To build a more intuitive picture of the orbits and the influence on their PS, in Fig. \ref{fig:knorbits} we show a 4$\times$4 diagram with increasing $\hat{a}$ from left to right and increasing $Q$ from top to bottom. Apparently, the effects of these two parameters on these orbits are in accord with those mentioned above.

\subsection{PS in Kerr-Sen spacetime}

The KS spacetime is described by the metric
\begin{align}\label{eq:ksmetric}
&\dd s^2=-\left({\frac{\Delta-\hat{a}^2\sin^2\theta}{\Sigma}}\right)\dd t^2 +{\frac{\Sigma}{\Delta}}d r^2\nn\\
&-{\frac{4\mu \hat{a} r\cosh^2\alpha\sin^2\theta}{\Sigma}}\dd t \dd\phi+\Sigma \dd \theta^2+{\frac{\Xi\sin^2\theta}{\Sigma}}\dd \phi^2,
    \end{align}
where
\begin{align}
&\Delta=r^2-2\mu r+\hat{a}^2,\nn\\
&\Sigma=r^2+\hat{a}^2\cos^2\theta+2\mu r\sinh^2\alpha,\nn\\
&\Xi=\left(r^2+2\mu r\sinh^2\alpha+\hat{a}^2\right)^2-\hat{a}^2\Delta\sin^2\theta,\nn\\
&\hat{a}=\frac{2J}{\mu (1+\cosh 2\alpha)},~\sinh^2\alpha=\frac{Q^2}{2M^2-Q^2},\nn\\
&\mu=M-\frac{Q^2}{2M} ,\nn
\end{align}
where $M,\,Q$ and $J$ are the mass, charge and spin angular momentum of the KS spacetime. Formally this spacetime has the same kind of parameters as the KN spacetime and when $Q=0$, it reduces to the Kerr spacetime. However, it has some features quite different from the KN spacetime. For example, the extreme KS BH happens when $Q^2=2M(M-\hat{a})$, which requires $0\leq \hat{a}\leq M$ and $0\leq Q\leq \sqrt{2}M$ \cite{Uniyal:2017yll}.

For the PS with circular approximation, we find that the equation determining $r_c$ once $(L,\,E)$ are fixed, is still a six-order polynomial whose solution is not feasible to write out here. Again, we assume that this $r_c$ is solved numerically and carry out the rest of the procedure.
At this $r_c$, the small quantity $\eta$ defined in Eq. \eqref{eq:etadef} for the KS spacetime becomes
\begin{align}
\eta_{\text{KS}}=&\left[\sqrt{\left(M r+Q^2\right)\left(\hat{a}^2 M+r \left(M (r-2 M)+Q^2\right)\right)}\right.\nn\\
&\left.\times\sqrt{\left(E^2-1\right)( M r+Q^2)+2 M^2}-2 \hat{a} E M^{5/2}\right]\nn\\
    &\left.\times\frac{1}{L \sqrt{M}\left(M (r-2 M)+Q^2\right)}\right|_{r=r_c}-1.
    \label{eq:etaks}
\end{align}
Expanding the metric functions at $r_c$, we find
\begin{subequations}
\begin{align}
A=&\frac{-M^2+Q^2+Mr_c}{Q^2+Mr_c}-\sum_{i=1}\frac{2M^3}{Q^4}\lb \frac{-Q^2r_c}{Mr_c+Q^2}\rb^{i+1}s^{i},\\
B=&-\frac{4\hat{a}M^2}{Q^2+Mr_c}-\sum_{i=1}\frac{4\hat{a}M^3}{Q^4}\lb \frac{-Q^2r_c}{Mr_c+Q^2}\rb^{i+1}s^{i}, \\
C=&\frac{r_c(Q^2+Mr_c)}{M}+\frac{\hat{a}^2(2M^2+Q^2+Mr_c)}{Q^2+Mr_c}\nn\\
&-\sum_{i=1}(-r_c)^{i+1}\left[ (i+1)r_c+\frac{Q^2}{M}\right.\nn\\
&\left.-\frac{2\hat{a}^2M^3}{Q^4}\lb \frac{Q^2}{Mr_c+Q^2}\rb^{i+1}\right]  s^{i},\\
D=&\frac{r_c(Q^2+Mr_c)}{\hat{a}^2M+Q^2r_c-2M^2r_c+Mr_c^2}\nn\\
&+\frac{2Mr_c^2(2M^2r_c^2-\hat{a}^2(Q^2+2Mr_c))}{(\hat{a}^2M+Q^2r_c-2M^2r_c+Mr_c^2)^2}s
+\calco(s)^2.
\end{align}
\end{subequations}
Substituting the coefficients of these expansions into Eq. \eqref{eq:yexp}, and further into Eq. \eqref{eq:deltaf1} together with $\eta$ in Eq. \eqref{eq:etaks}, we can then find the PS $\delta_{\text{SE,KS}}$ in the KS spacetime using the SE method. Since the $y_n$ are quite lengthy if no further limits are taken, we will not show their explicit forms here.
If we further take the large radius approximation, then as done in the Subsec. \ref{subsec:kn}, this shift can be expressed as series of $(p,\,e)$ too. Its form, as proven in Subsec. \ref{subsec:compmeth}, will be the same as the result using PN method.

For the PN$\Phi$ and PNU method, the asymptotic expansion of the metric \eqref{eq:ksmetric} is
\begin{subequations}
\begin{align}
A=&1+\frac{2M^2}{Q^2}\sum_{i=1}\lb -\frac{Q^2}{Mr}\rb^i,\\
B=&\frac{4\hat{a}M^2}{Q^2}\sum_{i=1}\lb -\frac{Q^2}{Mr}\rb^i,\\
C=&r^2+\frac{Q^2r}{M}+\hat{a}^2-\frac{2\hat{a}^2M^2}{Q^2}\sum_{i=1}\lb -\frac{Q^2}{Mr}\rb^i,\\
D=&1+\frac{2M}{r}+\frac{-\hat{a}^2+4M^2-2Q^2}{r^2}\nn\\
&+\frac{8M^4-(4M^2-2Q^2)(\hat{a}^2+2Q^2)}{Mr^3}+\calco\lb \frac{M}{r}\rb^4,
\end{align}
\end{subequations}
from which the coefficients are read off easily. Substituting these coefficients into Eq.
\eqref{eq:deltaf4}
%s. \eqref{eq:fexpcoeff} and \eqref{eq:elexpcoeff}, and then further into Eqs. \eqref{eq:deltaf2} and \eqref{eq:deltaf3},
we are able to find the PS in the KS spacetime as
\begin{align}
\delta_{\text{PN,KS}}=&\frac{\pi\left(6M-Q^2/M\right)}{p}-\frac{8\pi \hat{a}\sqrt{M}}{p^{\frac{3}{2}}} \nn\\
&+\frac{\pi}{8p^2}\left[12 \left(e^2+18\right) M^2+24\hat{a}^2\right.\nn\\
&\left.-4 \left(7 e^2+30\right) Q^2+3 \left(e^2+2\right) Q^4/M^2\right]\nn\\
&-\frac{2\pi \hat{a}\left[36M^{3/2}-(11+3e^2)Q^2/M^{1/2}\right]}{p^{\frac{5}{2}}}.
\label{eq:pnresks}
\end{align}
We note that both the leading order and the FD effect terms are the same as the KN spacetime and the deviation starts from the $p^2$ order. This implies that from the PS observation's point of view, it is not possible to distinguish the KN and KS spacetimes until we are able to measure the 2PN order in the future.

\begin{figure}[htp!]
\centering
\includegraphics[width=0.22\textwidth]
{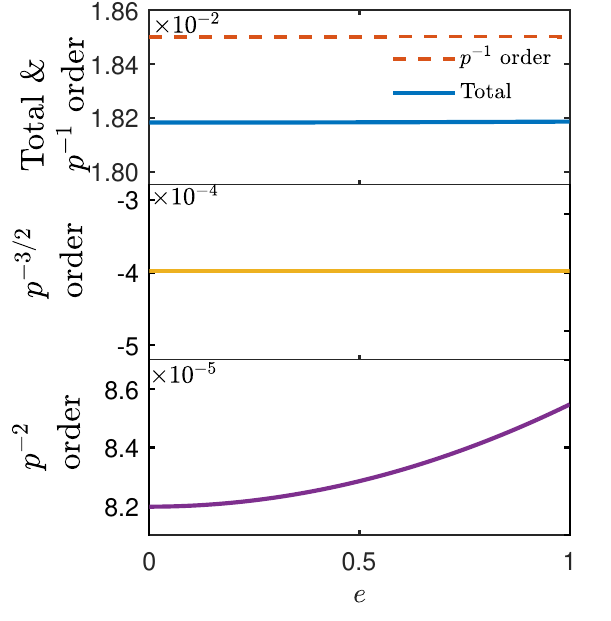}
\includegraphics[width=0.22\textwidth]
{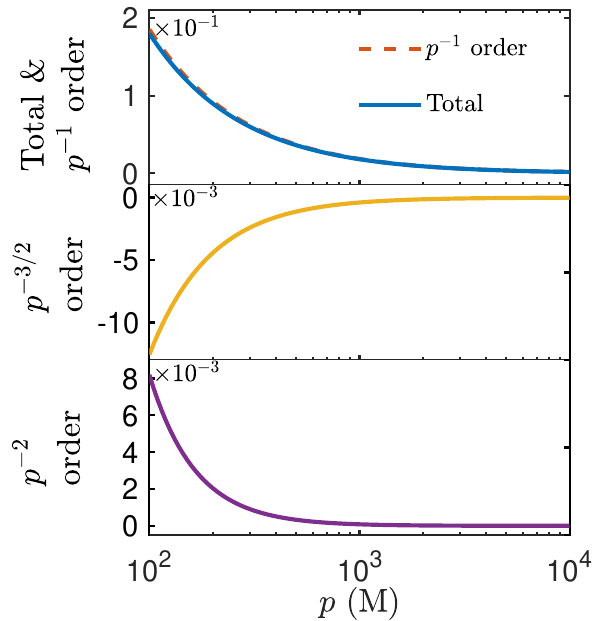}\\
(a)\hspace{3cm}(b)\\
\includegraphics[width=0.22\textwidth]
{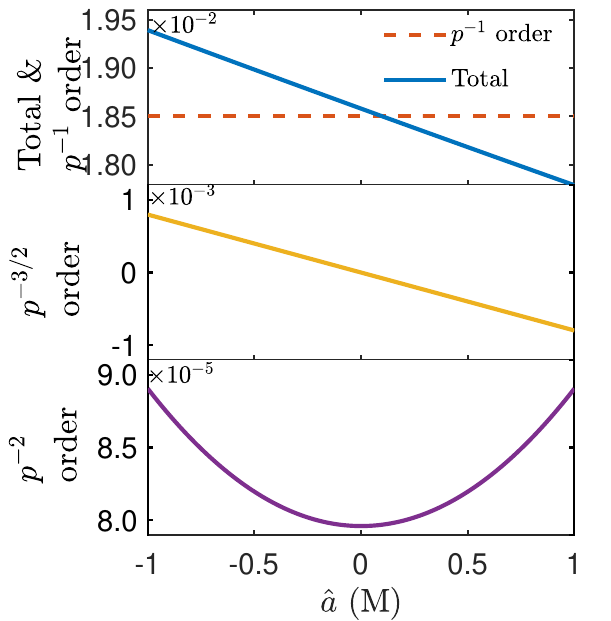}
\includegraphics[width=0.22\textwidth]
{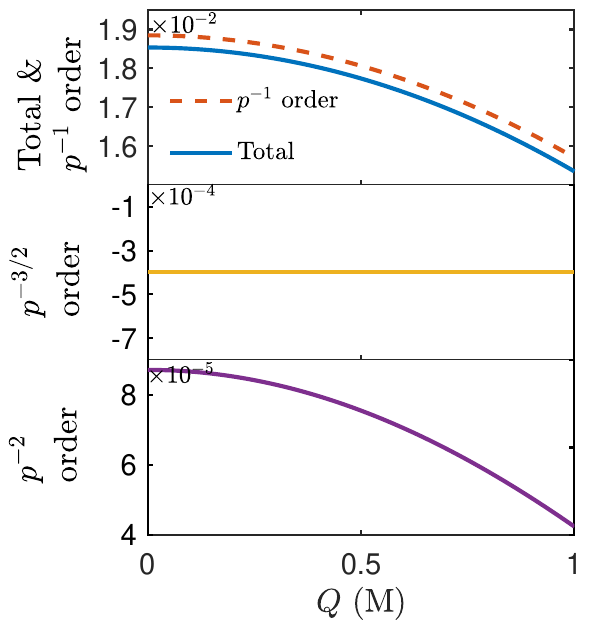}\\
(c)\hspace{3cm}(d)
\caption{
Dependence of the PS on $e$ (a), $p$ (b), $\hat{a}$ (c) and $Q$ (d) (top panel: total and leading order PS; middle panel: sub-leading PS; bottom panel: third order PS). The default parameters are $p=1000M,~e=1/100,~Q=M/3,~\hat{a}=M/2,~M=1$ except the one being varied.
}
   \label{fig:ks}
\end{figure}

We studied how accurate the KS spacetime PS, including the SE method result and Eq. \eqref{eq:pnresks}, are compared to PS obtained using direct numerical integration,
by plotting figures similar to Fig. \ref{fig:kn} (a)-(d). It is found that these results when the truncation order is high ($\bar{n}$=9) are as accurate as in Fig. \ref{fig:kn}. Therefore, we will only show the dependence of the total, leading, next-to-leading and third-order PS on parameters $e,~p,~\hat{a}$ and $Q$ in Fig. \ref{fig:ks} (a)-(d). The qualitative features in these plots are also very similar to those in Fig. \ref{fig:kn} because the metric of the KS spacetime is also highly similar to that of the KN. The main difference, which is still quantitative but not qualitative, happens in the bottom panel of Fig. \ref{fig:ks} (a) and (d) which corresponds to the $p^2$ order term in Eq. \eqref{eq:pnresks}: the slope that the order $p^{-2}$ term increases as $e$ increases (or $Q$ decreases) is slightly smaller (or larger) than those in Fig. \ref{fig:kn} (e) (or (h)).

\subsection{PS in KTN spacetime}

The line element of the KTN spacetime on the equatorial plane takes the form \cite{KTN1966}
\begin{align}
\dd s^2=&-\frac{r^2-2Mr-l^2}{r^2+l^2}\dd t^2-{\frac{4\hat{a}Mr+4l^2}{r^2+l^2}}\dd t \dd\phi\nn\\
    &+\frac{r^2+l^2}{r^2-2Mr+\hat{a}^2-l^2}\dd r^2\nn\\
    &+\left[r^2+l^2+\hat{a}^2+\frac{2\hat{a}^2(l^2+Mr)}{l^2+r^2}\right]\dd\phi^2,
    \label{eq:ktnmetric}
\end{align}
where $M,\,-l,\,\hat{a}$ are the spacetime mass,  gravitomagnetic monopole moment or the NUT charge and the spacetime spin parameter respectively \cite{Duztas:2017lxk}. The radius of this BH's horizons are $r_\pm =M\pm \sqrt{M^2-\hat{a}^+l^2}$ which requires $\hat{a}^2\leq M^2+l^2$ for its existence.

For the SE method, we first note that substituting the above metric into Eq. \eqref{eq:dlreq}, the resulting equation is an order ten polynomial of $r_c$, whose solution is still assumed to be obtained numerically. At this $r_c$, the quantity $\eta$ in the KTN spacetime is found from Eq. \eqref{eq:etadef} as
\begin{align}
&\eta_{\text{KTN}}=\left[\sqrt{\left(l^2+r^2\right)\left(\hat{a}^2 M+r \left(r (r-2 M)-l^2\right)\right)}\right.\nn\\
&\left.\times\sqrt{\left(E^2-1\right)( r^2+l^2)+2 Mr-2l^2}-2 \hat{a} E (l^2+Mr)\right]\nn\\
    &\left.\times\frac{1}{L \left(r (r-2 M)-l^2\right)}\right|_{r=r_c}-1.
    \label{eq:etaktn}
\end{align}
The metric functions \eqref{eq:ktnmetric} can be expanded at $r_c$ as
\begin{subequations}
\begin{align}
A=&\frac{-l^2-2Mr_c+r_c^2}{l^2+r_c^2}\nn\\
&-\frac{2r_c^2(2l^2r_c+M(r_c^2-l^2))}{(l^2+r_c^2)^2}s+\calco(s)^2,\\
B=&-\frac{4\hat{a}(l^2+Mr_c)}{l^2+r_c^2}\nn\\
&-\frac{4\hat{a}r_c^2(Mr_c^2+2l^2r_c-Ml^2)}{(l^2+r_c^2)^2}s+\calco(s)^2,\\
C=&l^2+r_c^2+\frac{\hat{a}^2(r_c^2+2Mr_c+3l^2)}{l^2+r_c^2}\nn\\
&-\left[ 2r_c^3-\frac{r_c^2(Mr_c^2+2l^2r_c-Ml^2)}{(l^2+r_c^2)^2}\right] s+\calco(s)^2,\\
D=&\frac{l^2+r_c^2}{r_c^2-2Mr_c+\hat{a}^2-l^2}\nn\\
&+\frac{r_c^2(Mr_c^2+(2l^2-\hat{a}^2)r_c-Ml^2)}{(r_c^2-2Mr_c+\hat{a}^2-l^2)^2}s+\calco(s)^2.
\end{align}
\end{subequations}

For the PN methods, we can work out the asymptotic expansion of the metric \eqref{eq:ktnmetric}
\begin{subequations}
\begin{align}
A=&1-\frac{2M}{r}-\frac{2l^2}{r^2}+\frac{2Ml^2}{r^3}+\calco\lb \frac{M}{r}\rb^4,\\
B=&-\frac{4\hat{a}M}{r}-\frac{4\hat{a}l^2}{r^2}+\frac{4\hat{a}Ml^2}{r^3}+\calco\lb \frac{M}{r}\rb^4,\\
C=&r^2+l^2+\hat{a}^2+\frac{2\hat{a}^2M}{r}+\frac{2\hat{a}^2l^2}{r^2}\nn\\
&-\frac{2\hat{a}^2Ml^2}{r^3}+\calco\lb \frac{M}{r}\rb^4,\\
D=&1+\frac{2M}{r}+\frac{-\hat{a}^2+4M^2+2l^2}{r^2}\nn\\
&+\frac{-4\hat{a}^2M+8M^3+6Ml^2}{r^3}+\calco\lb \frac{M}{r}\rb^4.
\end{align}
\end{subequations}
Note that the gravitomagnetic charge term has a negative sign, contrasting the plus sign of regular electric charge $Q$ or magnetic charge $P$ in RN metric with a magnetic charge. This also directly leads to its contribution to the PS with a different sign compared to the KN case (see Eq. \eqref{eq:pnresktn}).

Reading off the coefficients and substituting them into Eq. \eqref{eq:deltaf4},
%s. \eqref{eq:fexpcoeff} and \eqref{eq:elexpcoeff}, and then further into Eqs. \eqref{eq:deltaf2} and \eqref{eq:deltaf3},
the PS in the KTN spacetime is found as
\begin{align}
&\delta_{\text{PN,KTN}}=\frac{\pi\left(6M +2l^2/M\right)}{p}-\frac{8\pi \hat{a}\sqrt{M}}{p^{\frac{3}{2}}}\nn\\
&+\frac{\pi \left[6\hat{a}^2+3(18+e^2)M^2+(44+3e^2)l^2-l^4/M^2\right]}{2p^2}\nn\\
        &-\frac{8\pi \hat{a}(9M^{3/2}+4l^2/M^{1/2})}{p^{\frac{5}{2}}}. \label{eq:pnresktn}
\end{align}
Comparing to the KN result \eqref{eq:pnreskn}, we see that the contribution of $l^2$ at the leading order has an opposite sign to that of $Q^2$. This effectively changes the effect of the charge on the PS; that is, the larger the $l$, the larger the PS. The FD effect on the spacetime spin however is unaltered even quantitatively from the KN result. This term reduces after setting $e=0$ to the corresponding term in Ref. \cite{Chakraborty:2013naa} in the weak field limit.
If we replace $l^2$ by $-Q^2/2$, then we further notice that the $p^{-5/2}$ order term also coincides with that in the KN spacetime while the $p^{-2}$ order is still different. Therefore from the PS point of view, even if we leave the $l^2\leftrightarrow -Q^2/2$ difference out, the KTN spacetime is different from the KN starting from the $p^{-2}$ order.

\begin{figure}[htp!]
\centering
% \includegraphics[width=0.22\textwidth]
% {conkTN.pdf}
% \includegraphics[width=0.22\textwidth]
% {coaKTN.pdf}\\
\includegraphics[width=0.22\textwidth]
{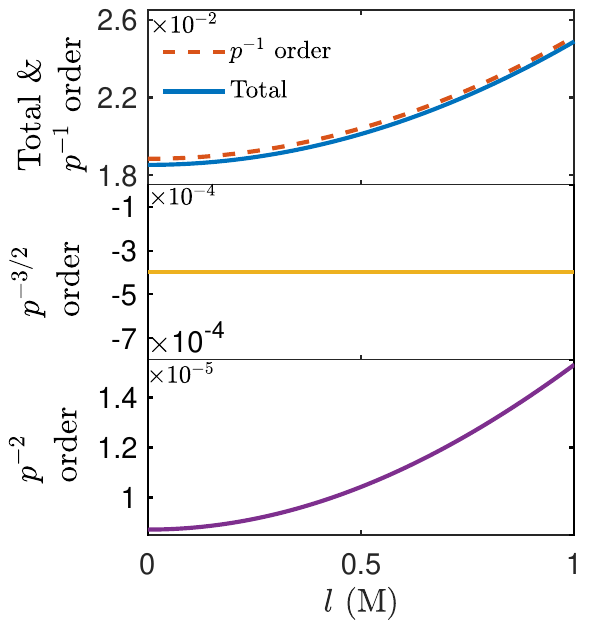}
\includegraphics[width=0.22\textwidth]
{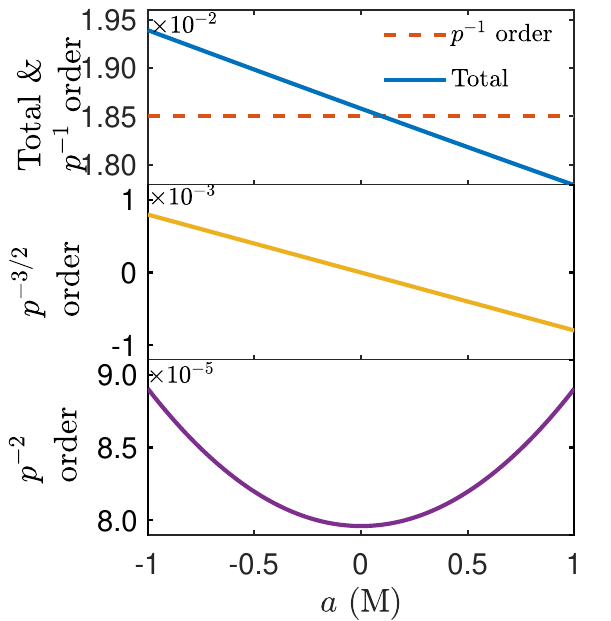}\\
(a)\hspace{3cm}
(b)\\
\includegraphics[width=0.55\textwidth]
{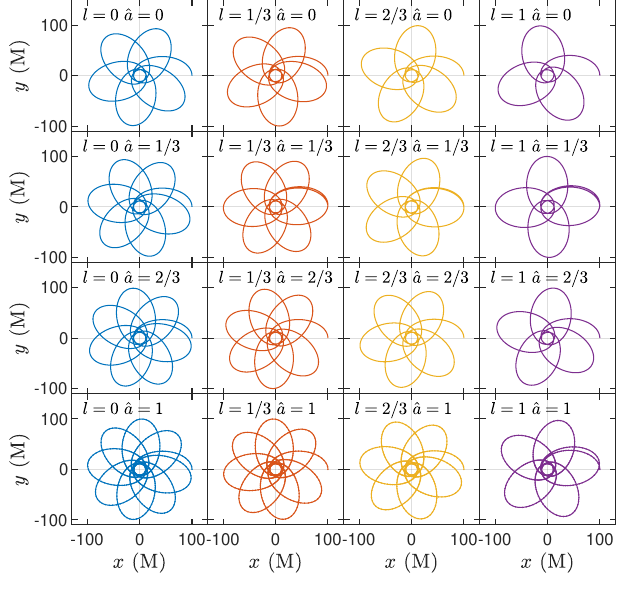}\\
(c)
\caption{Dependence of the PS on $Q$ (a), and $\hat{a}$ (b) (top panel: total and leading order PS; middle panel: sub-leading PS; bottom panel: third order PS). The default parameters in (a) and (b) are $p=1000M,~e=1/100,~l=M/3,~\hat{a}=M/2,~M=1$ except the one being varied. (c) Counter-clockwise rotating orbits in KTN spacetime with various values of $\hat{a}$ and $l$. The initial point is on the positive $x$-axis. Other parameters are $p=20M,~e=4/5,~M=1$. }
   \label{fig:ktn}
\end{figure}

In Fig. \ref{fig:ktn} (a) and (b), we plot the dependence of the total PS, and the leading order, next-to-leading order and third order contributions to the PS, on the parameters $l$ and $\hat{a}$. Their dependence on kinetic variables $p$ and $e$ are very similar to those in Figs. \ref{fig:kn} and \ref{fig:ks} and therefore not presented. It is seen that due to the opposite sign of the $l^2$ term in the leading order in Eq. \eqref{eq:pnresktn}, the leading and the third orders as well as the total PS increase as the NUT charge $l$ increases. This is the most significant difference from other charged spacetimes and might be used to constrain this parameter in observation. While for the spin parameter $\hat{a}$, its effect is similar to those in the KN (Fig. \ref{fig:kn} (g)) and KS (Fig. \ref{fig:ks} (c)) spacetimes both qualitatively and quantitatively.
In Fig. \ref{fig:ktn} (c) we show
the orbits in KTN spacetime with different NUT charge values and spacetime spin. From left to right in each row, it is seen that as the charge $l$ increases, the PS also increases, contrasting the case in Kerr spacetime in Fig. \ref{fig:knorbits}. From top to bottom in each column, the PS also increases as the increase of the corotating spin, which is qualitatively the same as in the Kerr case.

Finally, let us also mention that we also performed computation of the PS using the PN method for several other SAS spacetime, namely the rotating Ay\'on-Beato-Garcia, the rotating Bardeen and the rotating Hayward spacetimes, hoping to find more dramatic difference from the KN spacetime PS in the low orders. However it turned out that they differ from the KN case from at most $p^{-2}$ order. Since this order is not currently observable, we will not present a detailed analysis of them except to give their PS formulas in Appendix \ref{sec:apphigh}.
\section{Application to Astronomy \label{sec:appinastro}}

The PS data are available for a few objects/observations. The first is the PS observation for planets in the solar system, especially Mercury and other inner planets. The second is the precession of objects around Earth, mainly from the LAGEOS, LAGEOS II and Gravity Probe B experiments. The third is the more recently measured precession of S stars (S2) around the Sgr A*. In the following, we will attempt to apply our results, especially the KN result \eqref{eq:pnreskn} to these data.

\subsection{Precession around the Sun}

For Mercury's precession, it is known that there are many effects at various orders that contribute. Among them, the perturbations from Venus, Jupiter, Earth, Saturn and Mars contribute about 277.42, 153.99,  90.89, 7.32 and 2.48 ($^{\prime\prime}$/cty) respectively and the gravitoelectric (or Schwarzschild-like) effect from the Sun contributes about 42.98 ($^{\prime\prime}$/cty). Below this, the contribution from more outer planets Uranus and Neptune, as well as interaction among the planets, and then the solar oblateness contributes at the level of 0.01$\sim$0.1 ($^{\prime\prime}$/cty) \cite{park2017}.

From our Eq. \eqref{eq:pnreskn}, we see that beyond the Schwarzschild-like term, there is one more term due to the charge of the central body that is at the same order (order $p^{-1}$). Using $p=a(1-e^2)$
and dividing this term by the Mercury's orbital period $T=2\pi (a^3/M)^{1/2}$, it generate an extra rate of precession
\begin{align}
    \varpi_Q=\frac{Q^2}{2a^{5/2}M^{1/2}(1-e^2)}.
    \label{eq:vpiq}
\end{align}
Combining with the uncertainties in the measurement of the rate of precessions at this order, this extra term has been used to constrain the charge of the central body, such as that of the Sun and the Sgr A* in Ref. \cite{Iorio:2012dbo}.

The FD effect is even more insignificant compared to the effects mentioned above. For Mercury, it is expected to contribute only at the order of $\mathcal{O}(-0.0020)$  ($^{\prime\prime}$/cty) \cite{park2017}.
This effect is given by the $p^{-3/2}$ order term in Eq. \eqref{eq:pnreskn}. Again, converting it to the rate of precession, we find
\begin{align}
\varpi_J=-\frac{4J}{a^3(1-e^2)^{3/2}},
\label{eq:vpia}
\end{align}
where $J$ is the total angular momentum of the central body. Since other effects (asteroids, interaction between other planets etc.) are contributing to Mercury's precession at the same level, using the FD precession to constrain the solar angular momentum $J_\odot$ is not practical at present either.

If the precession of Mercury is eventually measured to even higher accuracy, then from Eq. \eqref{eq:pnreskn} we can estimate how precise the rate of precession has to be in order for the next order (order $p^{-2}$) to manifest.
Using $a=0.39$ (au) and $e=0.2056$ for Mercury and $M=1.99\times 10^{30}$ (kg) for solar mass, we can estimate that the $p^{-2}$ order rate of precession is
\begin{align}
    \varpi_{p^{-2}}\approx\frac{3(18+e^2)(GM_\odot)^{5/2}}{c^4a^{7/2}(1-e^2)^2}=2.01\times 10^{-5}~(^{\prime\prime}\text{/cty}).
\end{align}
This is currently about two orders smaller than the uncertainty in Mercury's rate of precession measurement \cite{park2017}.

If we assume that the central object carries a NUT charge $l$ as in the KTN spacetime, then the extra PS this charge causes will be governed by Eq. \eqref{eq:pnresktn}. At the leading order and switching to the rate of precession, this becomes
\begin{align}
    \varpi_l=-\frac{l^2}{a^{5/2}M^{1/2}(1-e^2)}.
    \label{eq:vpin}
\end{align}
Since Ref. \cite{Iorio:2012dbo} used the standard deviation of Mercury's orbit to estimate the order of the electric charge of the Sun, the sign and factor 2 difference between Eqs. \eqref{eq:vpiq} and \eqref{eq:vpin} will not matter if we use the same standard deviation to order estimate $l$. Therefore we can directly translate the constraint there $|Q_\odot|\lesssim 4.2\times 10^{17}$ (C) to that on the NUT/gravitomagnetic charge of the Sun, and obtain
\begin{align}
    |l_\odot| \lesssim 2.5\times 10^{25}~\text{A\,m}.
\end{align}

\subsection{Precession around the Earth}

Precessions of objects around Earth orbit provide another way to test gravitational theories. They have particular advantages in that such precessions can often be measured more precisely, and the fly orbit can be engineered. Such experiments include Gravity Probe B, the LAGEOS, LAGEOS II and LARES satellites. Here we will only study the results from the latter because they provide a stringent constraint than Gravity Probe B.

The LAGEOS, LAGEOS II \cite{lucchesi2011} and LARES satellites, together with GRACE data \cite{Lucchesi:2019kwm}, have been used to measure the FD effects of the earth. The quantity measured is the relative value of the secular shifts of the right ascension of the ascending node $\Omega$ or/and of the argument pericenter $\omega$, with respect to the GR predicted values. These values are found to deviate from the GR value of 1 in Ref. \cite{lucchesi2011,Lucchesi:2019kwm} by
\begin{align}
&\epsilon_{\omega}-1=(0.28\pm2.14)\times 10^{-3},\label{eq:epm1}\\
&\epsilon_{\Omega\omega}-1=(1.5\pm7.4)\times 10^{-3}\pm 16\times 10^{-3}.\nn
\end{align}
Here the first error budget is the true physical error within each routine of the measurement while the second is only a root-sum-square of a few results.

Several effects might contribute to these deviations, such as the ocean tide, other periodic effects, and de Sitter precession \cite{Lucchesi:2019kwm}. If we think of the above deviation to originate from the charge of the Earth, then using Eqs. \eqref{eq:vpiq}, \eqref{eq:vpia} and \eqref{eq:epm1} in the following
\begin{align}
    \varpi_Q\approx (\epsilon_\omega-1)\varpi_J
\end{align}
and considering data $a=1.22\times 10^7$ (meter) and $e=0.0044$ for LAGEOS \cite{Lucchesi:2019kwm} and $J_\oplus=5.86\times 10^{33}$ (kg\,m$^2$\,s$^{-1}$) for Earth \cite{Iorio:2023kmh}, we can  order estimate the charge of the Earth to be around
\begin{align}
    |Q_\oplus|\lesssim 2.9\times 10^{12}~\text{C}.
\end{align}
This is roughly one order better than the constraint in Ref. \cite{Iorio:2012dbo} which in turn is better than the lunar constraint.

If Earth carries NUT charge $l_\oplus$, then we can also use Eq. \eqref{eq:vpin} and result \eqref{eq:epm1} to estimate it. But this time the sign in \eqref{eq:vpin} matters and we can only use the lower boundary of the $1\sigma$ result, i.e., $\varepsilon_\omega-1\gtrsim -1.86\times 10^{-3}$. The corresponding constraint on $l_\oplus$ is
\begin{align}
    |l_\oplus|\lesssim 4.5\times 10^{20}~\text{A\,m}. \label{eq:nofearth}
\end{align}

\subsection{Precession around Sgr A*}

Compared to the precessions of the planets in the solar system or satellites around the Earth, precession around a BH apparently has its own advantages. The spin parameter $\hat{a}$ can reach $M$ so that the FD effect can be more prominent among various contributions. Moreover, the orbit around a BH can be much smaller relative to the central mass compared to the situations around the Sun or Earth and therefore the GR effects will be much more significant.

Recently, the precession of the star S2 around the Sgr A* SMBH has been detected
by the Gravity group \cite{GRAVITY:2020gka}. They measured that the total precession of S2 divided by the Schwarzschild-like precession (SP), i.e., the first term of Eq. \eqref{eq:pnreskn}, is about $f_{\text{SP}}\approx 1.10\pm 0.19$. The deviation of $f_{\text{SP}}$ from 1 can be used to constrain the BH parameters. Since this deviation 0.1 (the central value) is quite large, it can not be interpreted as due to the FD effect of the BH spin. Therefore the only possibility left is to use it to constrain the charge of the Sgr A* BH.

Dividing the SP term and the first charge term in Eq. \eqref{eq:pnreskn}, we find that
\begin{align}
    f_{\text{SP}}-1=0.1\approx \frac{Q^2}{6M^2}.
\end{align}
Using the fitted value of $M=4.261\times 10^6M_\odot$, the charge of the BH is constrained
\begin{align}
|Q_{\text{Sgr A*}}|\lesssim 5.7\times 10^{26}~\text{C}.
\end{align}
This is again about one order better than the value found in Ref.  \cite{Iorio:2012dbo} and similar to the values found in Ref. \cite{Ghosh:2022kit} ($5.5\times 10^{26} ~\text{C} -~8.9\times 10^{26} ~\text{C} $ and $1.3\times 10^{27} ~\text{C} -~1.4\times 10^{27} ~\text{C} $), although it seems this is still far from the theoretical constraint obtained in Ref.  \cite{Zajacek:2018ycb}.
Similar to the estimate of $l_\oplus$ in Eq. \eqref{eq:nofearth} for Earth, we can use the lower limit of the $1\sigma$ boundary of $f_{\text{SP}}$ to constrain the NUT charge of the Sgr A*. The result is found as
\begin{align}
    |l_{\text{Sgr A*}}|\lesssim 3.2\times 10^{34}~\text{A\,m}.
\end{align}

Since the spin of the BH is of interest to many other observables, such as the BH shadow size and shape,
while its value is not tightly constrained, we can estimate what precision should the measurement of S2 precession reach in order for the leading FD effect to be determined.
Using the $p^{-3/2}$ term in Eq. \eqref{eq:pnreskn} and data $e=0.88,~a=0.125^{\prime\prime}\times 8.247~\text{(kpc)}$ for S2 and $M=4.261\times 10^6M_\odot$ for Sgr A* \cite{GRAVITY:2020gka}, and assuming the extreme BH case, we find that the corresponding precession shift is  12.6$^{\prime\prime}$ per orbit period. In comparison, the Schwarzschild precession of S2 is about $12.1^\prime$ per orbit period \cite{GRAVITY:2020gka}.

\section{Conclusion and discussions \label{sec:concdis}}

We investigated in this work the PS of test particles in the equatorial plane of arbitrary SAS spacetimes.
Two perturbative methods, one fully general relativistic and the other post-Newtonian (including two variants) are developed. The SE method works for general radii but only small eccentricity $e$ while the PN methods work for arbitrary $e$ but only large semilatus rectum $p$.  All of these methods can be carried out to high orders and the results are shown to converge to true PS values.
Under limits of both large $p$ and small $e$, these results are shown to agree with each other analytically. The result in this case can be expressed as a dual power series of $p^{-1/2}$ and $e$, with coefficients determined as polynomials of the asymptotic expansion coefficients of the metric functions.

These PS results are applied to KN, KS and KTN spacetimes. Their results are used to analyze the correctness of the method, and more importantly the effects of the kinetic parameters $(p,\, e)$ and various spacetime parameters, including the spacetime spin $\hat{a}$ and charge $Q$. We paid special attention to the leading orders of $p$ (to $p^{-2}$) since they are most observational relevant to date. Generally, it is found that an electric charge would decrease the PS at the leading order (order $p^{-1}$) while a counter-rotating spin would increase it at order $p^{-3/2}$.

We then apply these PS to the observations of PS of planets around the Sun, man-made objects around the Earth and S star around the Sgr A* supermassive BH. Using relevant observational data, we contained the electromagnetic charge of the Earth and Sgr A* and discussed the required accuracy level for the 2PN order PS in Mercury's precession to be detected. Assuming KTN spacetime, the NUT/gravitomagnetic charge of the Sun, Earth and Sgr A* are constrained too. PN PS in the rotating ABG, rotating Bardeen and rotating Hayward spacetimes are supplemented in the appendix.

There are a few comments regarding this work that we would like to make. The first is that the current computation is under the {\it test particle} assumption, which ignores all gravitational wave radiation from the system and self-force from the smaller object. In other words, the PN results here are at the zeroth order of the small mass ratio in precessing binary problems, which have been very intensively studied in recent years. Comparing to PS obtained using more sophisticated PN methods (e.g. \cite{Damour:1988mr,Damour:1999cr,Capuzzo-Dolcetta:2023fhi}) in such situations, our method excels in its generality to arbitrary SAS spacetime and very high PN order but will work in larger mass-ratio systems. The second comment is regarding the exclusive applicability of the SE formula to available data. To our best knowledge, only for binary pulsars that are very close and accreting materials before falling into the BH, the orbit radius is small and the SE PS result has unique advantages. However in the former case, usually the mass ratio is not that extreme
\cite{Burgay:2003jj, Lyne:2004cj}
and in the latter case the viscosity can not be ignored. In other words, the test particle approximation is violated in these scenarios. The third comment is also somehow related to the test particle approximation. In the current work, we have ignored the extension and consequently the spin of the particle. However, as was known \cite{Akcay:2016dku,Kavanagh:2017wot,Iorio:2020tbe}, this can have a non-trivial effect on the orbital motion or PS of the particle. Therefore this will have to be taken into account when considering higher-order terms of the PS.

\acknowledgments

This work is partially supported by grants from NNSF and MOST China. The work of S. Xu is supported by the Wuhan University Students Innovation and Entrepreneurship Program.

\appendix

\section{Expansion of \texorpdfstring{$Y(\xi)$}{} in Eq. \eqref{eq:qphitransed}
\label{app:yhalfzero}}

The function $Y(\xi)$ in Eq. \eqref{eq:qphitransed} is defined as the sum of similar contributions from two branches
\begin{align}
    Y(\xi)=&\sum_{i=1,2}(-1)^{i+1}\left(-\frac{q^\prime_i(\xi)}{q_i(\xi)^2}\right)\nn\\
    &\times\sqrt{\frac{AD}{AC+B^2/4}} \frac{\beta}{\sqrt{\eta-\xi+2\beta}}.
\label{eq:yxidef}
\end{align}
In principle, using the Lagrange inverse theorem, we are only able to show that near an extreme point, the expansion of either $q_1(\xi)$ or $q_2(\xi)$ are half-of-integer power series of $\xi$ as shown in Eq. \eqref{eq:firstplus}. Naively one would expect that the such defined $Y(\xi)$ would also be a half-of-integer power series of $\xi$. However, because of the special property of the second-order extreme, we can have a stronger result: $Y(\xi)$ is actually a half-integer power series of $\xi$. We prove this in the following.

First, we show that the expansions of $q_1(\xi)$ and $q_2(\xi)$ have the same coefficients for their odd-order terms while opposite coefficients for their even-order terms. That is,
\begin{subequations}
    \label{eq:firstplus}
\begin{align}
q_2(\xi)=&\sum_{i=0}^\infty \lb  c_i\xi^i+d_i\xi^{(2i+1)/2}\rb,\label{eq:secondplus}\\
q_1(\xi)=&\sum_{i=0}^\infty \lb  c_i\xi^i-d_i\xi^{(2i+1)/2}\rb.
\label{eq:secondminus}\end{align}
\end{subequations}
From Eq. \eqref{eq:rtoxi} and that $r_c$ is a simple extreme of $h(r)$, we can know that we can expand the function $\xi=h(1/r)$ at $r_c$ as
\begin{align}
    \xi=h_2\lb \frac1r-\frac1{r_c}\rb^2+h_3\lb \frac1r-\frac1{r_c}\rb^3+\cdots.
\end{align}
Without losing generality we can assume that $h_2>0$ so that $\xi>0$.
Taking the square root of this equation, we can find that when $r<r_c$
\begin{align}
    \sqrt{\xi}=\lb \frac1r-\frac1{r_c}\rb\sqrt{h_2+h_3\lb \frac1r-\frac1{r_c}\rb +\cdots}
\label{eq:xipart1}
\end{align}
and when $r>r_c$
\begin{align}
-\sqrt{\xi}=\lb \frac1r-\frac1{r_c}\rb\sqrt{h_2+h_3\lb \frac1r-\frac1{r_c}\rb+\cdots}.
\label{eq:xipart2}
\end{align}
Now the right-hand sides of Eqs. \eqref{eq:xipart1} and \eqref{eq:xipart2} are the same and if we define it as function $g(r)$, i.e.,
\begin{align}
g(r)=\lb \frac1r-\frac1{r_c}\rb\sqrt{h_2+h_3\lb \frac1r-\frac1{r_c}\rb+\cdots},
\end{align}
then use the Lagrange inversion theorem, we know that there exists a unique inverse function $q(g)$ of $g(r)$ whose series form near $r=r_c$ is
\begin{align}
q(g)=\frac1{r_c}+k_1 g+k_2g^2+\cdots, \label{eq:qdefing}
\end{align}
where $k_1,~k_2,~\cdots$ are the expansion coefficients.
Now from Eqs. \eqref{eq:xipart1} and \eqref{eq:xipart2} we know that $g=\sqrt{\xi}$ for $r<r_c$ and then Eq. \eqref{eq:qdefing} yields
\begin{align} q(\sqrt{\xi})=\frac1{r_c}+k_1 \sqrt{\xi}+k_2\xi+\cdots
\end{align}
and $g=-\sqrt{\xi}$ for $r>r_c$ so that
\begin{align}
    q(-\sqrt{\xi})=\frac1{r_c}-k_1 \sqrt{\xi}+k_2\xi+\cdots.
\end{align}
The above two functions are nothing but the inverse functions $q_2(\xi)$ and $q_1(\xi)$ introduced in Sec. \ref{subsec:grse}. This proves the claim made at the beginning of this paragraph that the $q_1(\xi)$ and $q_2(\xi)$ take the form shown in Eq. \eqref{eq:firstplus}.

To proceed, let us call two series having the same coefficients for odd terms and opposite coefficients for even terms to be {\it even conjugate} series. Then the $q_1(\xi)$ and $q_2(\xi)$ in Eq. \eqref{eq:firstplus} are such even conjugate series. Instead, if the two series have the same coefficients for even terms but opposite coefficients for odd terms, then we can call them {\it odd conjugate}.
Then we claim, and it is not difficult to persuade oneself that the square, reciprocal and composite with the same regular function of two even conjugate series, still produce two even conjugate series; while the derivatives of two even conjugate series actually yield two odd conjugate series.
Based on these observations, we see that the two functions summed in Eq. \eqref{eq:yxidef} are even conjugate (note the extra minus sign).
Therefore when they are summed, the even terms from the two series cancel each other while the odd terms are doubled.
Since the leading two orders of $q_1(\xi)$ and $q_2(\xi)$ series are respectively $\xi^0$ and $\xi^{1/2}$, the $Y(\xi)$'s expansion should start from the $\xi^{-1/2}$ order and increase each time by one order. This is exactly the form we present in Eq. \eqref{eq:yexp}.

\section{PS in some spacetimes to higher orders\label{sec:apphigh}}

For future reference, here we present to PS in Kerr, RN and Schwarzschild spacetimes to higher orders, on top of those shown in Eq. \eqref{eq:pnreskn}.

For Kerr spacetime, the PS to order $p^{-9/2}$ is
\begin{align}
\delta_{\text{K}}=&\delta_{\text{KN}}\Big|_{Q=0}+ \frac{3 \pi  m \left[a^2 \left(50-6 e^2\right)+15 \left(e^2+6\right) m^2\right]}{2 p^3}\nn\\
&+\frac{6 \pi  a \sqrt{m} \left[2 a^2 \left(e^2-3\right)-9 \left(e^2+10\right) m^2\right]}{p^{7/2}}\nn\\
&+\pi  \left[72 a^4 \left(3-2 e^2\right)-60 a^2 \left(e^4+2 e^2-492\right) m^2\right.\nn\\
&\left.+105 \left(e^4+72 e^2+216\right) m^4\right]\frac{1}{32 p^4}\nn\\
&+\frac{2 \pi  a m^{3/2} \left[10 a^2 \left(6 e^2-43\right)-3 \left(173 e^2+630\right) m^2\right]}{p^{9/2}},
\end{align}
where $\delta_{\text{KN}}$ is given in Eq. \eqref{eq:pnreskn}.
The precession in RN spacetime to the $p^{-4}$ order is
\begin{align}
\delta_{\text{RN}}=&\delta_{\text{KN}}\Big|_{a=0}+ \frac{\pi}{8 m^3 p^3}  \left[180 \left(e^2+6\right) m^6\right.\nn\\
    &\left.-6 \left(19 e^2+126\right) m^4 Q^2-2 \left(e^2-31\right) m^2 Q^4-Q^6\right]\nn\\
    &+\frac{\pi}{64 m^4
   p^4} \left[210 \left(e^4+72 e^2+216\right) m^8\right.\nn\\
   &-180 \left(e^4+76 e^2+240\right) m^6 Q^2\nn\\
   &+6 \left(3 e^4+302 e^2+1428\right) m^4 Q^4\nn\\
    &\left.+20 \left(e^2-4\right) m^2 Q^6-5 Q^8\right].
\end{align}
In Schwarzschild spacetime, the PS to order $p^{-5}$ is
\begin{align}
\delta_{\text{S}}=&\delta_{\text{RN}}\Big|_{Q=0}
+
\frac{567 \pi  \left(5 \left(e^2+24\right) e^2+216\right) m^5}{32 p^5}\nn\\
&+\frac{231 \pi  \left(11664+9720e^2+810e^4+5e^6\right) m^6}{128 p^6}\nn\\
&+\frac{1287 \pi  \left(11664+13608e^2+1890e^4+35e^6\right) m^7}{128 p^7}
.\end{align}

In addition, we also computed the PN PS in the rotating Ay\'on-Beato-Garcia, the rotating Bardeen and the rotating Hayward spacetimes \cite{Abdujabbarov:2016hnw} to high orders. Here we also present these results for others' reference.
The line element of the rotating Ay\'on-Beato-Garcia spacetimes is given by
\begin{align}
    \dd s^2=&-f(r,\theta)\dd t^2-2\hat{a}\sin^2\theta (1-f(r,\theta))\dd t\dd \phi+\frac{\Sigma}{\Delta}\dd r^2\nn\\
    &+\Sigma \dd \theta^2+(\Sigma-\hat{a}^2f(r,\theta)-2)\sin^2\theta \dd \phi^2,
\end{align}
where $\Sigma$ is as in Eq. \eqref{eq:sigmadeltakndef} and
\begin{align}
&f(r,\theta)=1-\frac{2Mr\sqrt{\Sigma}}{(\Sigma+Q^2)^{\frac{3}{2}}}+\frac{Q^2\Sigma}{(\Sigma+Q^2)^2},\nn\\
&\Delta=f(r,\theta)\Sigma+\hat{a}^2\sin^2\theta.
\end{align}
Its PN PS to order $p^{-3/2}$ is found to be the same as corresponding orders of the KN result, and to order $p^{-5/2}$ we have
\begin{align}
&\delta_{\text{RABG}}=\delta_{\text{KN}}\Big|_{\leftarrow p^{-3/2}}+\frac{\pi}{p^2}[12\hat{a}^2+6(18+e^2)M^2\nn\\
    &-2(42+e^2)Q^2-Q^4/M^2]-\frac{8\pi \hat{a}(9M^2-2Q^2)}{\sqrt{M}p^{\frac{5}{2}}}
%\nn\\
%   &\pi\{180(6+e^2)M^6-6(238+43e^2)M^4Q^2\nn\\
%   &+4\hat{a}^2M^2[-6(-25+3e^2)M^2+(-11+e^2)Q^2]\nn\\
%   &+2(79+5e^2)M^2Q^4-Q^6\}\frac{1}{8M^3p^3}
.
\end{align}
The line element of the rotating Bardeen is
\begin{align}
    \dd s^2=&-(1-\frac{2mr}{\Sigma})\dd t^2-\frac{4\hat{a}mr\sin^2\theta}{\Sigma}\dd t\dd \phi+\frac{\Sigma}{\Delta}\dd r^2\nn\\
    &+\Sigma \dd \theta^2+(r^2+\hat{a}^2+\frac{2\hat{a}^2mr\sin^2\theta}{\Sigma})\sin^2\theta \dd \phi^2,
\end{align}
where $\Sigma$ and $\Delta$ are as in Eq. \eqref{eq:sigmadeltakndef} with $Q=0$ and
\begin{align}
m=M\left( \frac{r^2}{r^2+g^2}\right)^{\frac{3}{2}}.
\end{align}
The PN PS to order $p^{-3/2}$ is the same as the Kerr result, and to order $p^{-3}$ we have
\begin{align}
    \delta_{\text{RB}}=&\delta_{\text{K}}\Big|_{\leftarrow p^{-3/2}}+\frac{3\pi[2\hat{a}^2-6g^2+(18+e^2)M^2]}{2p^2}\nn\\
    &-\frac{72\pi\hat{a}M^{\frac{3}{2}}}{p^{\frac{5}{2}}}+3\pi M[15(6+e^2)M^2+\hat{a}^2(50-6e^2)\nn\\
    &-4(14+3e^2)g^2]\frac{1}{2p^3}.
\end{align}
And finally, for the rotating Hayward, its metric takes the form \cite{Abdujabbarov:2016hnw}
\begin{align}
    \dd s^2=&-(1-\frac{2mr}{\Sigma})\dd t^2-\frac{4\hat{a}mr\sin^2\theta}{\Sigma}\dd t\dd \phi+\frac{\Sigma}{\Delta}\dd r^2\nn\\
    &+\Sigma \dd \theta^2+(r^2+\hat{a}^2+\frac{2\hat{a}^2mr\sin^2\theta}{\Sigma})\sin^2\theta \dd \phi^2,
\end{align}
where
\begin{align}
m=M\frac{r^3}{r^3+g^3}.
\end{align}
And its PN PS to order $p^{-5/2}$ is also the same as Kerr, and to order $p^{-3}$ we have
\begin{align}
    \delta_{\text{RH}}=&\delta_{\text{K}}\Big|_{\leftarrow p^{-5/2}}+[-15(6+e^2)M^3+\hat{a}^2(-50+6e^2)M\nn\\
    &+2(4+e^2)g^3]\frac{3\pi}{2p^3}.
\end{align}

\end{document}